\documentclass[reprint,nofootinbib,amsmath,amssymb,fontenc aps]{revtex4-1}
\usepackage{graphicx,dcolumn,bm,hyperref,float}
\usepackage{anyfontsize}
\usepackage{color}

\newcommand{\be}{\begin{equation}}
\newcommand{\ee}{\end{equation}}
\newcommand{\beq}{\begin{equation}}
\newcommand{\eeq}{\end{equation}}
\newcommand{\ba}{\begin{eqnarray}}
\newcommand{\ea}{\end{eqnarray}}
\newcommand{\bea}{\begin{eqnarray}}
\newcommand{\eea}{\end{eqnarray}}
\newcommand{\bef}{\begin{figure}}
\newcommand{\eef}{\end{figure}}

\newcommand{\f}{\varphi}
\newcommand{\pf}{\varepsilon}
\newcommand{\pfl}{\psi}
\newcommand{\ac}{\alpha}
\newcommand{\bc}{\beta}
\newcommand{\acr}{\tilde{\alpha}}
\newcommand{\bcr}{\tilde{\beta}}
\newcommand{\rhob}{\sigma_\text{B}}
\newcommand{\Cfix}{\mathcal{C}}
\newcommand{\cb}{\gamma}
\newcommand{\mpl}{M_{\mbox{\tiny{Pl}}}}
\newcommand{\fu}{\mathcal{F}}
\newcommand{\argf}{\mathcal{Z}}
\newcommand{\pref}{\lambda}

\begin{document}

\title{Acausality in Superfluid Dark Matter and MOND-like Theories}

\author{Mark P.~Hertzberg}
\email{mark.hertzberg@tufts.edu}
\author{Jacob A.~Litterer}
\email{jacob.litterer@tufts.edu}
\author{Neil Shah}
\email{neil.shah@tufts.edu}
\affiliation{Institute of Cosmology, Department of Physics and Astronomy, Tufts University, Medford, MA 02155, USA
\looseness=-1}

\begin{abstract}
There has been much interest in novel models of dark matter that exhibit interesting behavior on galactic scales. A primary motivation is the observed Baryonic Tully-Fisher Relation in which the mass of galaxies increases as the quartic power of rotation speed. This scaling is not obviously accounted for by standard cold dark matter. This has prompted the development of dark matter models that exhibit some form of so-called MONDian phenomenology to account for this galactic scaling, while also recovering the success of cold dark matter on large scales. A beautiful example of this are the so-called superfluid dark matter models, in which a complex bosonic field undergoes spontaneous symmetry breaking on galactic scales, entering a superfluid phase with a 3/2 kinetic scaling in the low energy effective theory, that mediates a long-ranged MONDian force. In this work we examine the causality and locality properties of these and other related models. We show that the Lorentz invariant completions of the superfluid models exhibit high energy perturbations that violate global hyperbolicity of the equations of motion in the MOND regime and can be superluminal in other parts of phase space. We also examine a range of alternate models, finding that they also exhibit forms of non-locality. 
\end{abstract}

\maketitle

\tableofcontents

\section{Introduction}

Modern cosmology is built on a rather simple model of the universe: contents of baryonic matter (and radiation), a cosmological constant $\Lambda$, and cold dark matter, together known as $\Lambda$CDM, with dynamics controlled by general relativity. These pieces are highly compelling, since they are all allowed by general principles of (local) Lorentz symmetry and quantum mechanics at low energies. A range of observations of large scale structure and the detailed fluctuations of the CMB \cite{Planck2018} support this $\Lambda$CDM model, establishing its place as the dominant model of modern cosmology. Playing a critical role on large scales is cold dark matter, which forms the cosmic web, galactic halos, and so forth. Dark matter is a priori a very reasonable idea -- some new stable, massive particle beyond the Standard Model. Altogether the proposal of dark matter seems highly compelling.

\subsection{Galactic Tension}

While cosmological observations are in broad agreement with the $\Lambda$CDM model, there have been a number of claims of possible tension between $\Lambda$CDM and galactic scale observations. In particular, one of the strongest empirical correlations in extragalactic astronomy is the Baryonic Tully-Fisher Relation \cite{McGaugh:2000}, which relates the baryonic mass $M_b$ of galaxies to their asymptotic rotational velocities $v_r$ as $M_b \propto v_r^4$. However, it has been argued that a simple treatment of $\Lambda$CDM, namely in a simple collapse model, predicts that the scaling relation is different $M_b \propto v_r^3$ \cite{McGaugh:2011}. This discrepancy has been seen in some simulations as well \cite{Vogelsberger}. It is also the case that the observed relation has rather little scatter, while naive modeling of dark matter suggests significant scatter. At this stage it would be premature to conclude that this falsifies vanilla cold dark matter, as a full accounting for feedback effects from baryons, further numerical modeling, etc., may alter this conclusion. But the current tension is intriguing. 

Furthermore, there are possible other challenges to $\Lambda$CDM from observations of dwarf satellite galaxies in the local group which show that the most massive dark matter subhalos in our galaxy are too dense to host any of our most luminous dwarf satellites \cite{Boylan-Kolchin}. These dwarf satellites also have been observed to be co-rotating in planar formations that are difficult to explain using vanilla $\Lambda$CDM \cite{Pawlowski:2013kpa,Ibata}. In addition there are observed to be cores at the centers of galaxies, without an obviously compelling explanation \cite{deBlok:2009sp}.
Therefore there is some motivation to seek a new theory which alleviates the issues of $\Lambda$CDM on galactic scales. At this stage, the most reasonable null hypothesis is that $\Lambda$CDM will ultimately prevail, but it is worthwhile having an open mind.

A radical alternative theory to dark matter is to modify gravity on large scales to so-called MOdified Newtonian Dynamics (MOND) \cite{Milgrom:1983ca,Milgrom:1983pn,Milgrom:1983zz,Bekenstein:1984tv}. This postulates that at low velocities and weak gravitational fields, Newtonian gravity is replaced by a modified Newtonian force law: $F_G =m\, \mu(a/a_0) \,a$ where $F_G=GM_{enc}m/R^2$ is the standard gravitational force, and $\mu$ is a function of the acceleration. This function is assumed to have the asymptotic behavior of $\mu \rightarrow 1$ for large $a$, while $\mu \rightarrow a/a_0$ for small $a$ relevant to galactic halos.  This means that on galactic scales the acceleration is $a= \sqrt{a_0 G M_{enc}}/R$, rather than the usual $a=G M_{enc}/R^2$. Hence, for circular motion with $a=v_r^2/R$, one has $(v_r^2/R)^2=a_0GM_{enc}/R^2$, giving $M_{enc}\propto v_r^4$ matching the observed Baryonic Tully-Fisher Relation. The observed constant of proportionality is matched by taking $a_0\approx 10^{-8}$\,cm/s$^2$.

Such a model is entirely phenomenological, without any known microscopic construction. In fact, since general relativity is the unique local theory of massless spin 2 particles \cite{Weinberg:1964ew,Weinberg:1965rz,Deser:1969wk,Feynman,Khoury:2013oqa,Hertzberg:2016djj,Hertzberg:2017abn,Hertzberg:2017nzl,Hertzberg:2020yzl,Hertzberg:2020gxu}, one might think there cannot be any Lorentz invariant formulation of MOND. However, that is not true; by including one or more new scalar degrees of freedom, one can lift MOND to be Lorentz invariant in various ways that we will discuss (e.g., \cite{Bekenstein:2004ne,Bruneton:2008fk,Blanchet:2006yt,Bernard:2014psa,Skordis:2020eui}). One might also think that these extra degrees of freedom imply that one is necessarily including dark matter anyhow \cite{Calmet:2017voc}. However, that is also not true; in the most standard formulations of MOND (as opposed to the upcoming superfluid models we will analyze), the new scalar(s) are essentially always off-shell throughout the universe, acting as virtual force mediators, not dark matter. 

On the phenomenological level, while MOND has been successful on galactic scales at modeling rotation curves \cite{Sanders,Famaey:2011kh} and the aforementioned dwarf satellite structures \cite{Zhao}, it faces severe observational problems on cosmological scales. On large scales, the universe is almost homogeneous and isotropic, allowing for simple theoretical and semi-analytical work. Here it is found that while $\Lambda$CDM works remarkably well, the MONDian models fail to reproduce the observed CMB power spectra \cite{Skordis} or the observed fluctuations in the matter power spectra \cite{Dodelson}. There may be issues on smaller scales as well \cite{Zhao:2005xk}. 

\subsection{A Unified Framework}

Hence MOND and $\Lambda$CDM seem to perform well on different scales. An ideal solution may therefore be a kind of hybrid model which behaves like MOND on galactic scales, but asymptotes to CDM on cosmological scales. This idea was developed in recent years in the very interesting and novel work, known as SuperFluid Dark Matter (SFDM) \cite{Berezhiani:2015pia,Khoury, Khoury:Another, Khoury:USFDS}, and acted as motivation for the present study. The key idea is that the scalar(s) that are needed to be introduced to mediate a new gravitational force can also be on-shell and act as dark matter on cosmological scales. On galactic scales, it undergoes a phase transition to a new superfluid phase, with a novel scaling in its effective theory, allowing it to mediate a type of MONDian force between baryons. The superfluid phonons mediate a MONDian force by coupling directly to the baryon density $\rhob$. Specifically, the idea is to build a theory with some new massive complex scalar dark matter $\Phi$. 

On large scales, its relevant Lagrangian is that of a regular massive scalar minimally coupled to gravity as (signature $-+++$)
\beq
\mathcal{L}_\text{SFDM}=
\sqrt{-g}\,\left[-{1\over2}g^{\mu\nu}\partial_\mu\Phi\partial_\nu\Phi^*-{1\over2}m^2|\Phi|^2+\ldots\right]
\eeq
ensuring that it reproduces the successes of $\Lambda$CDM on cosmological scales. On galactic scales, one introduces some non-trivial dynamics that causes the field to organize into a condensate. Then one decomposes the field in terms of its magnitude and phase as $\Phi=\rho\,e^{i(\theta+mt)}$, where the phase $\theta=\theta({\bf x},t)$ is the Goldstone from the spontaneous breaking of a phase symmetry. The dynamics need to be of a very special variety so that the effective non-relativistic Lagrangian of the Goldstone is of the form:
\begin{equation}
	\mathcal{L}_\text{SFDM} = 
	\ac\,Y \sqrt{|Y|} + \bc\, \theta \,\rhob+\ldots\label{SFDMlow}
\end{equation}
where $\ac$, $\bc$ are constants, $\rhob$ is the baryonic mass density, and
\beq
Y = \dot{\theta} - m\,\phi_N - \frac{1}{2m} (\nabla \theta)^2
\eeq
is a special combination of field derivatives and the Newtonian gravitational potential $\phi_N$.

The reason this type of low energy Lagrangian is pivotal to get the requisite MONDian dynamics can be explained as follows: Consider the static limit and let us ignore ordinary gravity for simplicity. Then $Y\approx-(\nabla\theta)^2/(2m)$ (plus a small chemical potential correction from $\dot\theta$), which is evidently negative; we shall return to this point later in the paper. The Lagrangian is then $\mathcal{L}_\text{SFDM}\approx -\ac\,((\nabla\theta)^2/(2m))^{3/2}+\bc\,\theta\,\rhob$, and the Euler-Lagrange equation for $\theta$ is
\beq
-{3\ac\over(2m)^{3/2}}\nabla\!\cdot\!(\nabla\theta|\nabla\theta|)=\bc\,\rhob
\eeq
The nonrelativistic equation of motion for a baryon subject to this fifth force is that its acceleration is ${\bf a}=\bc\,\nabla\theta$. For a spherically symmetric mass density $\rhob$, we can solve this pair of equations to obtain
\beq
{\bf a}=-\mbox{sign}(\ac)\sqrt{|\bc|^3(2m)^{3/2}M_{enc}\over12\pi |\ac|}\,{\hat{r}\over R} \label{accel}
\eeq
Hence so long as the coefficient $\ac$ of the fractional power in Eq.~(\ref{SFDMlow}) is positive, this force produces attraction, with magnitude of acceleration of $a=\sqrt{a_0GM_{enc}}/R$, with $a_0\sim|\bc|^3m^{3/2}/(\ac\,G)$. This matches the desired MONDian behavior, as described above, by choosing the quantity $|\bc|^3m^{3/2}/\ac$ accordingly. 

Later we shall recapitulate concrete Lorentz invariant models in the literature that achieve all these desired features (as well as develop some variations). Altogether this appears to be very promising.

 \subsection{Outline of this Work}
 
The form of the low energy Lagrangian Eq.~(\ref{SFDMlow}) features some peculiarities, in particular the fractional power of the kinetic term $Y$ in $\mathcal{L}_\text{SFDM}$. In this work, we are driven by the question if any Lorentz invariant version of the above SFDM construction exists that avoids pathologies and that, furthermore, could arise from a sensible UV completion. In fact there is a long history of addressing this kind of question on the space of effective theories that can have a possible UV completion; for example, see Ref.~\cite{Aharonov:1969vu}. If some version of string-based conjectures are correct, then most effective theories cannot be realized within string theory. But here we focus on a much more well established consistency criteria: we would like to know whether there are signs of inconsistency from causality breakdown within the effective theory itself. This would mean they cannot have a UV completion in the usual sense of Wilsonian effective field theory. Such an imposition makes sense without direct appeal to any string-based conjectures, but are reinforced if the conjectures are true. 

As a well known example, Lorentz invariant effective theories that at first sight appear local, can sometimes turn out to exhibit some form of non-locality upon closer examination. In particular, when certain inequalities are violated on the coefficients of higher dimension operators, the effective theory can exhibit superluminality around non-trivial backgrounds, and cannot possess a regular Lorentz invariant UV completion \cite{Arkani-Hamed}. Perhaps even more seriously is the direct breakdown of the Cauchy initial value problem.

In this paper we will focus on a set of causality and locality constraints on a family of Lagrangians involving various powers of some relativistic kinetic term $X$, so-called K-essence theories, as these are relevant to the various superfluid models, and related models, that are able to achieve some of the desired galactic phenomenology. For related work on using constraints on pure MOND models, see Refs.~\cite{Bruneton:2006gf,Bruneton:2007si}. The basic idea is as follows: Suppose a scalar field $\varphi$ sets up some condensate background $\varphi_b$, such as that from a superfluid. In that case, high energy scalar particles will propagate along an effective metric $G^{\mu \nu}_\varphi$ that we can derive from the K-essence Lagrangian. This propagation maintains a basic notion of causality if this effective metric is globally hyperbolic \cite{Waldtextbook}, otherwise catastrophic instabilities ensue and/or the initial value problem is compromised. A necessary condition of hyperbolicity is that the signature of $G^{\mu \nu}_\varphi$ must match the Lorentzian signature of the spacetime metric. We will utilize this constraint, as well as related constraints from subluminality and cluster decomposition, to examine the causal structure of a range of theories including $\mathcal{L}_\text{SFDM}$. While there has been previous work on identifying forms of superluminality in pure MOND models, our work here on the general breakdown of hyperbolicity in SFDM, plus results in other models, is new.

This paper is organized as follows:
In Section \ref{Constraints}, we develop causality constraints for a class of K-essence models.
In Section \ref{Examples} we show how theories with a known UV completion obey these conditions.
In Section \ref{Simple}, we present some very simplistic Lorentz invariant models that exhibit MONDian behavior.
In Section \ref{Superfluid}, we describe the basic dark matter superfluid model and show explicitly how it behaves as MOND on small scales and asymptotes to CDM on large scales.
In Section \ref{SuperfluidTest}, we extend the constraints we developed in Section \ref{Constraints} to a model with a $U(1)$ symmetry, which we then apply to the relativistic superfluid Lagrangian, proving general results.
In Section \ref{Other} we examine alternative formulations of SFDM. 
Finally, in Section \ref{Conclusions}, we conclude.

\section{Causality Constraints on Field Theories}\label{Constraints}

Let us consider a class of so-called K-essence models (e.g., see \cite{Garriga:1999vw,ArmendarizPicon:1999rj,Rendall:2005fv}), in which the kinetic term for some scalar field can be a generic function, i.e., an action of the form
\beq
S = -\int d^4 x\sqrt{-g}\left[F(X,\varphi)\right] \label{kessence}
\eeq
where 
\beq
X\equiv{1\over 2}g^{\mu\nu}\partial_\mu\f\partial_\nu\f
\eeq
We will work in the $-+++$ signature. The classical equations of motion that follow from this action are
\bea
&&\left[ F'(X,\f) g^{\mu\nu}+F''(X,\f)\partial^\mu\f\partial^\nu\f \right] \partial_\mu\partial_\nu\f \nonumber\\
&&=-2X{\partial F'\over\partial\f}+{\partial F\over\partial\f} \label{CEOM}
\eea
where primes mean derivatives with respect to $X$, and the spacetime derivatives can be lifted to covariant derivatives in a curved spacetime. One may also include a potential for the field, which will be unimportant to our discussion.

Now in order to access the causal structure of this classical equation of motion, let us suppose that the field has some background solution $\f_b$, which may be a function of space and/or time. Then let us study a small perturbation $\pf$ around this background solution as
\beq
\f=\f_b+\pf\label{pert}
\eeq
Even though $\f_b$ may depend on space and/or time, we are interested in perturbations $\pf$ that are much more rapidly changing. This is an essential diagnostic to learn about the causal structure of the theory, since signal speeds are associated with the high energy limit of particles on top of the background in a Lorentz invariant theory. Note that this is to be contrasted to the low energy collective phonon excitations, which propagate at the sound speed. In Ref.~\cite{Mistele} the focus was on ensuring the sound speeds were subluminal, but this is only one requirement for consistency. The more general requirement is that high energy particles exhibit subluminality too; we shall return to this discussion in Section \ref{TwoField}. 

When we substitute Eq.$\,$(\ref{pert}) into Eq.$\,$(\ref{CEOM}) and work to $\mathcal{O}(\pf)$ we only keep terms that involve {\em two} derivatives acting on $\pf$, since first derivative or zero derivative terms are subleading in this limit.  This leads to the equation for high energy perturbations
\beq
G^{\mu\nu}_\f\partial_\mu\partial_\nu\pf=0
\label{Geff}\eeq
where we have defined
\beq
G^{\mu\nu}_\f = F'(X_b,\f_b) g^{\mu\nu}+F''(X_b,\f_b)\partial^\mu\f_b\partial^\nu\f_b
\label{GeffDef}\eeq
From the point of view of the high energy perturbation $\pf$, the tensor $G^{\mu\nu}_\f$ is playing the role of an effective background metric. 

In order to ensure that perturbations on this effective metric exhibit standard causal evolution, one demands that the metric maintains the same $-+++$ signature. One can verify that two of the eigenvalues of $G^{\mu\nu}_\f$ are both $F'$, hence these must both be positive. In addition, one finds that the determinant of the metric is 
\beq
\mbox{Det}[G^{\mu\nu}_\f]=-F'(X,\f)^3(F'(X,\f)+2X F''(X,\f))
\eeq
which needs to be negative. Altogether, in order to ensure global hyperbolicity, we need the $F$ function to obey the following pair of conditions
\begin{align}
(A)\,\,\,\,\, A &\equiv F'>0 \label{A}\\
(B)\,\,\,\,\, B &\equiv F'+2X F''>0 \label{B}
\end{align}
which is in agreement with the work of Ref.~\cite{Bruneton:2006gf} (also, see Ref.~\cite{Babichev:2007dw}). Relatedly, the corresponding Hamiltonian exhibits a derivative instability if either of these conditions are violated. If the background mainly depends on time, then the instability will be in the temporal direction, indicating a ghost. 

Another important condition for a theory to possess a sensible Lorentz invariant UV completion is that it avoids  superluminal signal propagation. To assess this, one first divides the equation of motion by $F'(X_b,\varphi_b)$ to obtain a more regular looking wave equation.
Now to illustrate the basic idea, consider a background $\varphi_b$ that mainly depends on time $t$. Ultimately, it is important that it also have some spatial dependence in order to produce a localized region that asymptotes to the vacuum at infinity; this allows one to compare signal speed through the medium $\varphi_b$ to light speed in vacuum. But here we shall use a mainly time dependent background to illustrate the idea; the spatially dependent case can be easily derived too, with the same final result. On the other hand, we allow high energy perturbations to depend on both space and time as usual. Furthermore, let us focus on flat space $g^{\mu\nu}=\eta^{\mu\nu}$. Then Eqs.~(\ref{Geff},\,\ref{GeffDef}) become
\beq
-\left(1-{F''\over F'}\dot\varphi_b^2\right)\ddot\pf+\nabla^2\pf=0
\eeq
Thus we see that the speed of propagation of the perturbation is $c_{\mbox{\tiny{signal}}}^2=1/(1-(F''/F')\dot\varphi_b^2$). Now since we already need to remain in the regime in which $F'>0$ in order to ensure hyperbolicity, and we have $\dot\varphi_b^2>0$, then the signal speed will be superluminal, $c_{\mbox{\tiny{signal}}}^2>1$, if $F''>0$. Hence the condition for subluminality is
\begin{align}
(C)\,\,\,\,\, C &\equiv -F''\geq 0 \label{C}
\end{align}
We note that if one of  (A) or (B) are violated, then there is no regular wave propagation, and so this signal speed is no longer directly meaningful. However, if {\em both} (A) and (B) are violated, then there is once again a notion of wave propagation as the entire metric has flipped sign. In this case, the condition for subluminality is that condition (C) is {\em also} violated; we shall return to this situation at the end of Section \ref{XSquaredTerm}. However, if (A) and (B) are both violated, the corresponding Hamiltonian would pick up an overall minus sign, which means a potential instability when coupling to matter.

In the following sections, we will examine these conditions (A), (B), (C) for a variety of real scalar field theories. Collectively, we shall refer to any such violation as a form of ``acausality"; we are therefore using this as a generic name to include, superluminality, breakdown of standard hyperbolicity, and ghost behavior. We will generalize these conditions to a complex field in Section \ref{ComplexCausality} and then examine them in detail in the SFDM models. We emphasize that these are not merely ``strong coupling" problems; they represent real breakdown in aspects of the causal behavior of the purported Lorentz invariant theories. They strongly suggest that the effective theories are not applicable in the relevant regimes of interest.

\section{Examples of Consistent Causality}\label{Examples}

\subsection{Canonical Action}

As a first very simple example, consider a theory with a canonical kinetic term and potential
\beq
F_{can} = X + V(\varphi) \label{canonical}
\eeq
Then we trivially satisfy the above conditions. This represents the standard action for a scalar. This can of course play the role of cold dark matter, but does not obviously give rise to the correct galactic behavior.

\subsection{DBI Models}

As another example, consider the rather more non-trivial action for the so-called DBI models (arising from extra dimensional models) which have been put forth as models of inflation \cite{Silverstein:2003hf,Alishahiha:2004eh}
\beq
F_{DBI}=2T\sqrt{1+X/T} +V(\varphi)
\label{DBI}\eeq
where $T$ is related to the tension and is allowed to be a function of $\varphi$, i.e., $T=T(\varphi)$, or just a constant. Note that in the small energy density regime $X\ll T$, this simplifies to $F_{DBI}\approx F_{can}$, and so is standard. However for  values of $X/T$, the departures are significant. This model is not especially relevant to the problem of galactic behavior; we are simply including it here as a concrete example of a K-essence Lagrangian. Let us compute the above values of $A$, $B$, and $C$. We find
\bea
&&A={1\over\sqrt{1+X/T}}>0\\
&&B = {1\over(1+X/T)^{3/2}}>0 \\
&&C = {1\over 2T(1+X/T)^{3/2}} > 0
\eea
As indicated, $A$, $B$, and $C$ are evidently positive, satisfying the above constraints. Since the DBI model has been argued to have an embedding in string theory, and therefore has a UV completion, it was to be expected that it obeys all these causality conditions.

In this model, one might be concerned that the field evolves to obtain $1+X/T<0$. However, this is not actually a concern. To see this, we can compute the energy density, by forming the Hamiltonian density $\mathcal{H}$. In flat space, it is easy to show
\beq
\mathcal{H}={2T+(\nabla\varphi)^2\over\sqrt{1+X/T}} +V(\varphi)
\eeq
This shows it is unlikely for $1+X/T$ to pass through zero under time evolution, because then the energy density is blowing up, which would likely increase the integrated energy and the energy would not be conserved. Hence the phase space can be self-consistently restricted to the regime $1+X/T>0$, and conditions (A), (B), and (C) are always satisfied.

\subsection{Perturbative Theories}

Often when one is exploring an effective field theory, one stays within the domain in which the higher dimension operators are small compared to the leading order operators. Of course this is required for their contribution to the scattering of elementary particles, for otherwise one would likely be beyond the cutoff of the effective theory. However, this is not a general rule when expanding around some condensate background (as an example, inflation often involves super-Planckian field values, even though it is a sensible effective field theory). For the upcoming superfluid models, we will in fact need to enter the regime in which the condensate background allows the higher dimension operators to be as large, or larger, than the leading operators. 

In any case, for the present discussion let us restrict our attention to the perturbative regime, where higher dimension operators are small. We shall consider a pure kinetic model, with a tower of operators of the form
\beq
F=X +\sum_{n=n_{min}}g_n \,X^n/\Lambda^{4(n-1)} 
\eeq
where $g_n$ is a coefficient of the $n^{th}$ term, with $n$ an integer starting at $n_{min}\geq2$, and $\Lambda$ is some mass scale that sets the standard cut off on high energy scattering. The leading term $X$ represents the usual kinetic term, while the sum represents interaction terms in a theory with a shift symmetry $\varphi\to\varphi+\varphi_0$. One could also include other terms, such as Galileons, which we shall return to briefly in Section \ref{Galileon}; but we will not focus on those terms in the discussion here. The natural value of $n$ for the leading interaction is $n_{min}=2$. However, as we shall discuss in the upcoming superfluid models, it will be important to also consider a special case where $g_2=0$ or small and the leading behavior is instead provided by $n=3$, or higher. 

Working perturbatively and tracking only the leading order contributions we have
\bea
&&A\approx1>0\\
&&B \approx1 > 0 \\ 
&&C = -\sum_{n=n_{min}}n(n-1) g_n\, X^{n-2}/\Lambda^{4(n-1)}
\eea
So by working perturbatively, the hyperbolicity conditions (A) and (B) are trivially satisfied because the kinetic structure is dominated by the canonical kinetic term. 

For the subluminality condition (C), we see that if the $g_2$ term is significant, then we immediately obtain the condition 
$g_2\leq0$, which is the familiar sign theorem of Ref.~\cite{Arkani-Hamed} (taking into account $\Delta\mathcal{L}=-F$). If, on the other hand, the $g_2$ term is small, we can also turn our attention to the $g_3$ term. Together they say that the subluminality condition is 
\beq
-g_2-3\,g_3\,X/\Lambda^4\geq0
\eeq
For $g_3>0$ (which will occur in the upcoming superfluid dark matter models), this condition puts a bound on the value of $X$ to be 
$X\leq -g_2\,\Lambda^4/(3\,g_3)$. For $g_2=0$, this would restrict $X\leq0$, which would be immediately violated by almost all solutions. For 
$g_2<0$, but small, this puts a potentially tight bound on the regime of validity of the effective theory to avoid superluminality. It is therefore an important question whether such an effective theory is useful; we shall return to related issues when we discuss the superfluid model in Section \ref{XSquaredTerm}.

\section{Simplistic Lorentz Invariant MOND Models}\label{Simple}

\subsection{Non-Smooth Lagrangians}\label{NonSmooth}

Arguably, the simplest and most naive way to build a Lorentz invariant theory that achieves the MONDian dynamics is simply to start with the desired low energy Lagrangian Eq.~(\ref{SFDMlow}) and by brute force lift it to be Lorentz invariant by replacing $Y\to X$ and $\rhob\to T_\text{B}$, where $T_\text{B}$ is some Lorentz invariant source, such as the trace of the baryonic energy-momentum tensor. Hence the simplest incarnation of a Lorentz invariant theory with some kind of MONDian dynamics is that the scalar field $\varphi$ has Lagrangian
\beq
\mathcal{L} = -\acr\,X\sqrt{|X|} - \bcr\,\varphi\,T_\text{B}+\ldots
\label{nonsmth}\eeq
Recall that $X=g^{\mu\nu}\partial_\mu\varphi\partial_\nu\varphi$/2, hence in the limit in which we ignore gravity, it is $X=(-\dot\varphi^2+(\nabla\varphi)^2)/2$, and so in the static limit it is $X=(\nabla\varphi)^2/2$. This means $X\propto-Y$ (up to small corrections). Furthermore, since $T_\text{B}=-\rhob$ in this limit, we recover the static limit of Eq.~(\ref{SFDMlow}), by re-scaling $\acr$ and $\bcr$ appropriately. This indicates the MONDian dynamics follows.

While this formally achieves both MONDian dynamics within a Lorentz invariant theory, this Lagrangian is clearly highly peculiar. A fractional power in a Lorentz invariant theory is very unusual. In the quantum theory, this does not possess a standard expansion in terms of creation and annihilation operators that act on particle states, which fundamentally is what defines quantum fields. Correspondingly, there will presumably not be a usual form of cluster decomposition in order to maintain locality \cite{Weinberg}. In addition, we know that on large scales, this MONDian dynamics performs poorly observationally, as we outlined in the introduction. 

We will fix all these problems shortly by turning to the novel superfluid dark matter models in the following sections. However, for now, let us proceed and test the causal structure of this theory against the above causality conditions. It is simple to check that with the above action $F=\acr\,X\sqrt{|X|}$, the conditions for hyperbolicity are
\bea
&&A = {3\,\acr\over2}\sqrt{|X|}>0\\
&&B = 3\,\acr\sqrt{|X|} > 0
\eea
Since we know $\acr>0$, as required to achieve MONDian dynamics as in Eq.$\,$(\ref{accel}), these first two constraints are satisfied. Finally, the quantity relevant to signal speed propagation is
\beq
C = -{3\,\acr\,X\over4 |X|^{3/2}}
\eeq
In order to remain subluminal, $C\geq0$, we therefore need $X\leq0$. Since $X=g^{\mu\nu}\partial_\mu\varphi\partial_\nu\varphi$ is a kinetic term, one normally cannot restrict its range of values, so this condition would normally be violated. Furthermore, in the MONDian regime, we have $X\approx(\nabla\varphi)^2/2>0$, meaning that the condition is definitely violated in this regime. So although this above simple Lorentz invariant model obeys global hyperbolicity, it permits superluminality around non-trivial backgrounds. 

\subsection{Smooth Lagrangians}\label{Smooth}

On the other hand, studying the above theory perturbatively is not particularly meaningful, because it does not have a regular kinetic term and so does not behave well in the standard vacuum. In this case, any conclusions about superluminality can be taken with a note of caution. In any case, the fractional power in the Lagrangian means this simple theory fails all sorts of standard tests for sensible quantum field theories, such as cluster decomposition. One can improve the situation by lifting the action to various smooth forms. An example that we can consider is
\beq
F={X+\acr\,X^3/\Lambda^{6}\over(1+X^6/\Lambda^{24})^{1/4}}+{1\over2}m^2\varphi^2
\label{smth}\eeq
which now has a well defined expansion around the vacuum.
This recovers the above action in the $\Lambda\to0$ limit. At small energy densities, this theory recovers the canonical form for a massive scalar $F=X+m^2\varphi^2/2$, while at large energy densities it asymptotes to $F=\acr\,X\sqrt{|X|}+m^2\varphi^2/2$. The kinetic term appears to give the desired MONDian phenomenology. 
Some related examples are given in Ref.~\cite{Khoury:2014tka}.

We will not repeat it, but it should be clear that subluminality again breaks down here in the MONDian regime. Furthermore, this theory can also violate hyperbolicity, depending on the value of $\acr\,\Lambda^2$. We find that this hyperbolicity problem can be avoided for all $X$, so long as the following inequality is obeyed
\bea
\acr\,\Lambda^2>0.1041\,\,\,\,\,\mbox{to obey (A)}\\
\acr\,\Lambda^2>1.6191\,\,\,\,\,\mbox{to obey (B)}
\eea

Now, since this theory possesses a power series expansion around the vacuum $X=0$, we can treat this as a kind of effective theory. However, when operating in the high energy density regime relevant to MOND, $X\gg\Lambda^4$, one may be beyond the cut off of the effective theory, thus entering a non-unitary regime. So this is not clearly a well behaved model in the usual sense of effective field theory. It would be advantageous to have a model that could avoid this potential breakdown in unitarity. 

Moreover, there is a pressing phenomenological problem with this theory, and that is related to the mass $m$. In order for the MONDian force to be long-ranged, one would require $m$ to be extremely small, presumably no larger than $1/L_{gal}$, where $L_{gal}$ is the size of a galaxy. However, this also presents a potential problem if this is to simultaneously act as dark matter on large scales, because various analyses, such as Lyman $\alpha$ forest, etc., suggest that the dark matter mass cannot be so tiny \cite{Boyarsky:2008xj,Viel:2005qj,Viel:2013fqw,Schutz:2020jox,Benito:2020avv}. This motivates one to find another model in which the mass $m$ can be appreciable in order to act as cold dark matter on large scales, and yet simultaneously mediate a long-range MONDian force despite this mass.
This leads us to the discussion of the SFDM models in the next section.

\section{Superfluid Model}\label{Superfluid}

A useful idea would be to introduce a massive complex field that undergoes spontaneous symmetry breaking, with a massless Goldstone that can mediate the desired long-range force. This leads us to the novel superfluid dark matter models from the literature, as we now describe.

\subsection{Complex Field}

In the very interesting Ref.~\cite{Khoury}, a model of dark matter was introduced that exhibits MOND behavior on galactic scales, while relaxing to standard CDM behavior on cosmological scales. This model involves a complex scalar field $\Phi$. 
A fully relativistic version of the theory is given by
\beq
F = {1\over2}\left(X+m^2|\Phi|^2\right)
+{\Lambda^4\over6(\Lambda_c^2+|\Phi|^2)^6}\left(X+m^2|\Phi|^2\right)^3\,\,\,\,\,\,\label{SuperL}
\eeq
where we have updated the definition of $X$ to that which is appropriate to a complex field with a $U(1)$ symmetry, namely
\beq
X=g^{\mu\nu}\partial_\mu\Phi\partial_\nu\Phi^*
\eeq
In addition there are assumed to be terms that couple $\Phi$ to baryons, such as $\Delta\mathcal{L}=-\cb\,\Lambda\,\theta\, T_b/\mpl$, where $T_b$ is the energy-momentum tensor of the baryons, $\theta$ is the phase of $\Phi$ (see ahead to Eq.~(\ref{decompose})), and $\cb$ is a dimensionless coupling. This is essential for $\theta$ to mediate a new long-ranged force. Such a term introduces a small explicit breaking of the $U(1)$ symmetry, which is a shift symmetry in $\theta$, although it may be technically natural. 

A curious feature of the above Lagrangian is that the same mass $m$ appears both in the first term and the cubic term. This appears fine-tuned from the top down point of view, although somehow natural from the bottom up point of view as it will lead to a scaling regime. In any case, we will not address this issue further here.

Note that this form of the action avoids any fractional powers, so it may be thought of as some kind of effective theory, and furthermore we can enter an interesting scaling regime for $|\Phi|\gg\Lambda_c$. 
This action is missing a natural operator, which is the quadratic power $\propto X^2$. However, we will reinstate this later in Section \ref{XSquaredTerm}, finding qualitatively similar results. Then in Section \ref{GenAn} we will generalize this to all possible models with the desired scaling.

The CDM regime occurs for small values of $\Phi$ and $X$. In this regime, the action simplifies to 
\beq
F \approx {1\over2}\left(X+m^2|\Phi|^2\right)\,\,\,\,\,\,\,\,(\mbox{CDM regime})
\eeq
which is a Lagrangian for a standard massive scalar, and therefore behaves as CDM (at least for length scales larger than its de Broglie wavelength). It is of the form of Eq.$\,$(\ref{canonical}) and to this leading order analysis would appear to obey conditions (A--C); however, we shall return to this issue later.

\subsection{Phase Transition}

If we were to treat the higher dimension operators in Eq.$\,$(\ref{SuperL}) perturbatively, then the hyperbolicity conditions (A) and (B) would still be satisfied because the extra terms would be small in this regime (although the subluminality condition (C) would still be a potential issue). But this is {\em not} the MOND regime. The point of the extra terms is that they become as important as the other terms in some new regime, giving rise to the MOND phenomenology. In such a regime, it is less clear that the hyperbolicity conditions are obeyed.

In this other regime, we decompose the field into modulus $\rho$ and phase $\theta$ as
\beq
\Phi = \rho\,e^{i(\theta+mt)}
\label{decompose}\eeq
The extraction of the phase factor $mt$ in the exponent allows for an identification of the non-relativistic regime, since the fast oscillations are extracted, and then we can assume that the remaining functions $\rho$ and $\theta$ are slowly varying in space and time. 

In order to determine the resulting low energy effective action, it is useful to express $X$ in terms of $\rho$ and $\theta$. By writing the metric in the nonrelativistic limit as 
\beq
ds^2=-(1+2\phi_N)dt^2+dx^2+dy^2+dz^2
\eeq
and working only to linear order in $\phi_N$, dropping all two-derivatives terms, since the fields $\rho,\,\theta$ are slowly varying in the low energy regime, it is readily found the following relation
\beq
X+m^2|\Phi|^2=(\nabla\rho)^2-2\,m\,\rho^2\,Y
\label{XY}
\eeq
where 
\beq
Y\equiv\dot\theta-m\,\phi_N-{(\nabla\theta)^2\over 2m}\label{Ydef}
\eeq
which is the same function we reported earlier in the introduction. One can check that this expansion is valid for $\Lambda_c^2<\Lambda\,m$.

In the MOND regime, the field $\rho$ is heavy and can be integrated out. At tree-level (classical) it can be shown that it is linked to the value of the phase field to leading order in a gradient expansion as
\beq
\rho^2 = \Lambda\sqrt{2m|Y|}\label{rhoSoln}
\eeq
Then by self-consistently dropping spatial gradients acting on $\rho$, we can insert Eq.~(\ref{rhoSoln}) into eqs.~(\ref{XY}) for $X$ and then into Eq.~(\ref{SuperL}) to obtain the effective action
\beq
F=-{2\Lambda(2m)^{3/2}\over3}Y\sqrt{|Y|}\,\,\,\,\,\,\,\,(\mbox{MOND regime})
\label{F3}\eeq
which achieves the desired kinetic part of the MONDian low energy action Eq.~(\ref{SFDMlow}) with $\ac=2\Lambda(2m)^{3/2}/3$ and $\bc=\cb\,\Lambda/\mpl$. The corresponding value for the turnover acceleration is $a_0\sim\cb^3\Lambda^2/\mpl$. This is all in agreement with the work of Ref.~\cite{Khoury}. 

One can check that in the MOND regime it is indeed self-consistent to ignore the gradient terms acting on $\rho$ as follows: As we mentioned earlier, the acceleration is related to the gradient of $\theta$ by $\nabla\theta={\bf a}/\bc\sim\sqrt{\bc\, m^{3/2}M_{enc}/\ac}\,\hat{r}/R\sim\sqrt{\cb M_{enc}/\mpl}\,\hat{r}/R$. Hence we can estimate $Y$ as
\beq
Y\approx-{(\nabla\theta)^2\over 2m}\sim -{\cb\,M_{enc}\over\mpl\,m\,R^2}
\eeq
So if we compare the gradient term to the non-gradient term in Eq.~(\ref{XY}) we have
\beq
{(\nabla\rho)^2\over 2m\,\rho^2 \,Y}\sim{1\over m R^2Y}\sim-{\mpl\over\cb\,M_{enc}}
\label{ratio}\eeq
For any reasonable value of the dimensionless coupling $\cb$, this ratio is minuscule because there is a factor of the mass of the galaxy on the denominator. Hence the strategy of ignoring spatial derivatives of $\rho$ is indeed justified in the MONDian regime.

As an aside, we note that this model predicts the polytropic equation of state $P\propto\rho^3$, which does not accurately describe the cores of galaxies, as shown in Ref.~\cite{Deng:2018jjz}. But we shall not focus on that issue further here.

Importantly, we see that this clever idea is achieving the MONDian dynamics from a reasonable type of Lorentz invariant starting point, along with CDM on large scales, rather than the simplistic fractional power starting point of Section \ref{NonSmooth}, or even the smooth version of Section \ref{Smooth} which didn't explain why there is a mode light enough to mediate a long-ranged force.

\section{Test for Causality in Superfluid Model}\label{SuperfluidTest}

\subsection{Causality Constraints in Models with $U(1)$ Symmetry}\label{ComplexCausality}

In this case we have a complex scalar. This can be broken up into a pair of real scalars $\phi_j$, with $j=1,2$, as
\beq
\Phi=(\phi_1+i\,\phi_2)/\sqrt{2}
\eeq
We can then decompose $X=X_1+X_2$, with $X_j={1\over2}(\partial\phi_j)^2$ (with this definition, the leading order kinetic and mass terms in Eq.~(\ref{SuperL}) will have a factor of $1/4$ when written in terms of these real fields, which is unconventional). Each scalar obeys the full relativistic equation of motion
\bea
&&\sum_{j=1}^2\left[{\partial F\over\partial X_j}\eta^{\mu\nu}\delta^{ij}+{\partial^2F\over\partial X_i\partial X_j}\partial^\mu\phi_i\partial^\nu\phi_j\right]\!\partial_\mu\partial_\nu\phi_j\nonumber\\
&&=-\sum_{j=1}^2{\partial^2 F\over\partial X_i\partial\phi_j}\partial^\mu\phi_i\partial_\mu\phi_j+{\partial F\over\partial\phi_i}
\eea
The $U(1)$ symmetry of the model ensures that ${\partial F/\partial X_j}=F'$ and ${\partial^2 F/\partial X_i\partial X_j}=F''$. Hence this simplifies to 
\bea
&&\sum_{j=1}^2\left[F'\eta^{\mu\nu}\delta^{ij}+F''\partial^\mu\phi_i\partial^\nu\phi_j\right]\partial_\mu\partial_\nu\phi_j\nonumber\\
&&=-\sum_{j=1}^2{\partial F'\over\partial\phi_j}\partial^\mu\phi_i\partial_\mu\phi_j+{\partial F\over\partial\phi_i}
\eea
which generalizes Eq.$\,$(\ref{CEOM}) to a two-field model organized by a $U(1)$ symmetry. 

We again decompose the field into a background piece $\phi_j^b$ and a high energy perturbation $\pf_j$ as
\beq
\phi_j=\phi_j^b+\pf_j
\eeq
The linear equation for $\pf_j$ is easily obtained by keeping only the terms that involve two-derivatives on $\pf_j$ 
\beq
\sum_{j=1}^2\left[F'\eta^{\mu\nu}\delta^{ij}+F''\partial^\mu\phi_i^b\partial^\nu\phi_j^b\right]\partial_\mu\partial_\nu\pf_j=0
\label{ej}\eeq
(with $F'$ and $F''$ both evaluated on the background values). To solve a set of linear differential equations, one would normally take a Fourier transform. However, the dependence on space and time in the background $\phi_i^b$ complicates this. Nevertheless, we can proceed by again using the fact that $\phi_i^b$ varies slowly, while $\pf_j$ varies rapidly. To proceed lets identify the normal modes by decomposing $\pf_j$ as an approximate plane wave as
\beq
\pf_j(t,{\bf x})=\tilde{\pf}_j(t,{\bf x})\,e^{i\,k_\mu(t,{\bf x})\,x^\mu}+c.c
\eeq
where $\tilde{\pf}_i$ and $k_\mu$ are assumed to be slowly varying. By inserting this in to Eq.~(\ref{ej}), we self consistently only allow derivatives to act on the $x^\mu$ in the exponent and ignore the other derivatives; this is usually known as the ``geometric optics" limit. Since there are a pair of fields, this becomes a $2\times2$ matrix problem
\beq
\!\!\!\left(\!\!\begin{array}{cc}
F'k^2 + F''(\partial^\mu\phi_1^bk_\mu)^2 & F''\partial^\mu\phi_1^bk_\mu\partial^\nu\phi_2^bk_\nu\\
F''\partial^\mu\phi_1^bk_\mu\partial^\nu\phi_2^bk_\nu & F'k^2 + F''(\partial^\mu\phi_2^bk_\mu)^2
\end{array}\!\!\right)\!
\left(\!\begin{array}{c}\tilde{\pf}_1\\\tilde{\pf}_2
\end{array}\!\right)=0
\eeq
where $k^2=\eta^{\mu\nu}k_\mu k_\nu$.  In order for this to possess nontrivial solutions, we require the determinant of the matrix to vanish. 
The determinant is easily found to be
\beq
F'k^2[(F'\eta^{\mu\nu}+F''(\partial^\mu\phi_1^b\partial^\nu\phi_1^b+\partial^\mu\phi_2^b\partial^\nu\phi_2^b))k_\mu k_\nu]=0
\eeq
This gives the dispersion relations for each of the two normal modes of the system. The first normal mode arises from setting the first factor here to zero, i.e., $k^2=0$. This is the standard relation for a massless degree of freedom. We recover this here since the theory has a Goldstone which moves on a null light ray.

More interesting is the second normal mode, which arises from setting the term in the square brackets $[\ldots]$ to zero. If we form the  linear combination 
\beq
\pfl=\partial^\mu\phi_1^b\partial_\mu\pf_1+\partial^\mu\phi_2^b\partial_\mu\pf_2
\eeq
then we obtain the normal coordinate for this second mode. It obeys the corresponding differential equation
\beq
G^{\mu\nu}_{\phi}\partial_\mu\partial_\nu\pfl=0
\eeq
where we have again introduced a kind of effective metric
\beq
G^{\mu\nu}_\phi = F' g^{\mu\nu}+F''(\partial^\mu\phi_1^b\partial^\nu\phi_1^b+\partial^\mu\phi_2^b\partial^\nu\phi_2^b)
\eeq
This generalizes eqs.~(\ref{Geff},\,\ref{GeffDef}) to the second normal mode of this two-field model.
The corresponding hyperbolicity and subluminality conditions are then immediately the same conditions (A--C) in eqs.~(\ref{A},\,\ref{B},\,\ref{C}), where the prime now means a derivative with respect to the full $X$.

\subsection{Application to Superfluid Model}\label{SuperfluidTestB}

Let us take the derivatives that are relevant to the tests for consistency with causality. To express the results for $A,\,B,\,C$ it is useful to rescale them as
\beq
A={\tilde{A}\over\zeta(\Phi)},\,\,\,\,\,B={\tilde{B}\over\zeta(\Phi)},\,\,\,\,\,C={\tilde{C}\over\zeta(\Phi)}
\eeq
where $\zeta(\Phi)\equiv(\Lambda_c^2+|\Phi|^2)^6$ is manifestly positive. So checking on the signs of $A,\,B,\,C$ is equivalent to checking on the signs of the numerators $\tilde{A},\,\tilde{B},\,\tilde{C}$. These numerators are given by
\bea
\tilde{A}&=&(\Lambda_c^2+|\Phi|^2)^6/2+\Lambda^4(X+m^2|\Phi|^2)^2/2\label{tA}\\
\tilde{C}&=& -\Lambda^4(X+m^2|\Phi|^2)\label{tC}
\eea
with $\tilde{B}$ conveniently given in terms of these as
\beq
\tilde{B}=\tilde{A}-2X\tilde{C}\label{tB}
\eeq
Clearly we have $A>0$, so condition (A) is satisfied. However, there is ambiguity in the signs of $B$ and $C$, so it is unclear whether the other two conditions are satisfied. Nevertheless, this can readily be determined, as follows.

Since we wish to take the non-relativistic limit, it is important to note that factors like
\beq
\dot\theta,\,\,\,{(\nabla\theta)^2\over2m}
\eeq
are both relevant and both contribute to $Y$. To make this manifest, one can rescale spatial co-ordinates as
${\bf x}={\bf x}'/\sqrt{m}$,
and then the above pair of terms scale as
$\dot\theta,\,\,\,(\nabla'\theta)^2/2$.
So even in the non-relativistic limit, where $m$ is large compared to the spatial variation in these fields, both of these terms can play a role.

When we are in the MONDian regime, we can use the smallness of $(\nabla\rho)^2$ as established in Eq.~(\ref{ratio}) to simplify Eq.~(\ref{XY}) to
\beq
X+m^2|\Phi|^2\approx-2\,m\,\rho^2\,Y
\label{Xapprox}\eeq
This allows one to readily estimate $\tilde{C}$. Furthermore, one can check that the $F'$ contributions to $B$ are sub-dominant to the $2XF''$ contributions in this regime. Hence we easily obtain $\tilde{B}\approx-2X\tilde{C}$. Then to the most leading order, we have the even simpler estimate $X\approx -m^2\rho^2$. 
Putting this altogether, we find that in the MONDian regime we have
\bea
\tilde{B} &=& 4\, m^3\,\Lambda^4\rho^4\,Y \label{Bval} \\
\tilde{C} &=& 2\, m\,\Lambda^4\rho^2\,Y \label{Cval}
\eea
where $\rho$ is given by Eq.~(\ref{rhoSoln}).
Hence we see that the signs of both $B$ and $C$ are determined by the sign of $Y$. But this is a serious problem, as we explain: Recall that we earlier said that in the MONDian regime, we have
\beq
Y\approx-{(\nabla\theta)^2\over2m}
\eeq
which is evidently negative (there can be a chemical potential contribution too, but the MOND regime still corresponds to $Y<0$). 
This ensures that $B$ is negative. Thus the relativistic perturbations about the superfluid model of Eq.$\,$(\ref{SuperL}) fails the condition of hyperbolicity. Since the background mainly depends on time due to the fast varying $e^{imt}$ factor in Eq.~(\ref{decompose}), this hyperbolicity breakdown is in the temporal direction, meaning the theory carries a ghost. 

Since $C$ is also negative, it suggests a potential breaking of subluminality also. However, once hyperbolicity is broken there is no longer a meaningful signal speed that can be defined. Nevertheless, we can also identify a breakdown of superluminality in the usual sense: Suppose we have yet to fully enter the above MOND regime, but are in the transition region from CDM towards the MOND regime. When we are deeply in the CDM regime, we have $B\approx F'\approx 1/2>0$, and hyperbolicity is obeyed, and we have regular wave propagation. However there can be regions of phase space in which $C=-F''$ is still negative, so there will be superluminality. For example, consider a semi-relativistic region of space in which $X=-|\dot\Phi|^2+|\nabla\Phi|^2$ happens to be small, then $B\approx F'$ can be positive, while $C$ can be negative. Altogether, in some intermediate regime with $C<0$ and superluminality, or deeply in the MOND regime with $B<0$ and hyperbolicity breakdown, this is all in tension with standard notions of causal propagation and consistency.

This is in contrast to some of the earlier examples, such as the DBI models, which obey the hyperbolicity and subluminality conditions at every point in space and time in all of phase space. It is also in contrast to the earlier examples of ``simplistic Lorentz invariant MOND models" of Section \ref{Simple} which obey the hyperbolicity conditions (although they disobeyed the subluminality condition).

\subsection{Including Quadratic Term}\label{XSquaredTerm}

In the above model, there is an omission of another natural operator, namely a quadratic term $(X+m^2|\Phi|^2)^2$. For completeness, let us also include this term. It can potentially be quite important from the point of view of causality, because if it enters the Lagrangian with a positive coefficient, then it can help to avoid problems of superluminality \cite{Arkani-Hamed}. This was also included briefly in the work of Ref.~\cite{Khoury}. The full action is
\bea
F &=& {1\over2}\left(X+m^2|\Phi|^2\right)+{\Lambda^4\over6(\Lambda_c^2+|\Phi|^2)^6}\left(X+m^2|\Phi|^2\right)^3
\nonumber\\
&&-{g\,\Lambda^2\over2(\Lambda_c^2+|\Phi|^2)^3}\left(X+m^2|\Phi|^2\right)^2\label{SuperLg}
\eea
where $g$ is a new coefficient that sets the size of this new term. We now find that the coefficients $\tilde{A},\,\tilde{C}$ controlling the causality constraints are altered from Eqs.~(\ref{tA},\,\ref{tC}) to
\bea
\tilde{A}&=&(\Lambda_c^2+|\Phi|^2)^6/2+\Lambda^4(X+m^2|\Phi|^2)^2/2\nonumber\\
&&-g\,\Lambda^2(\Lambda_c^2+|\Phi|^2)^3(X+m^2|\Phi|^2)\\
\tilde{C}&=&-\Lambda^4(X+m^2|\Phi|^2)+g\,\Lambda^2(\Lambda_c^2+|\Phi|^2)^3
\eea
with Eq.~(\ref{tB}) still applicable to obtain $\tilde{B}$.
So by choosing $g>0$ one can hope to satisfy the second and third conditions as it may help to ensure that $\tilde{C}>0$. On the other hand, $g>0$ may introduce a problem with the first condition. As we shall see, in the superfluid regime, no value of $g$ will succeed in avoiding the causality problems.

By again decomposing the field $\Phi$ in terms of modulus $\rho$ and phase $\theta$ in the superfluid regime, one can again solve for $\rho$ to leading order. The result in Eq.~(\ref{rhoSoln}) can be found to be altered to\footnote{In Ref.~\cite{Khoury} there appears to be a sign error in the first term under the square root in their Eq.~(96).}
\beq
\rho^2 = \Lambda\sqrt{g\,m\,Y+\sqrt{4+g^2}\,\,m\,|Y|}
\eeq
Note that the argument under the square root is always positive for any value of $g$ or $Y$, both positive or negative; so this is well defined. The corresponding values of $\tilde{A},\,\tilde{B},\,\tilde{C}$ are found to be
\bea
\tilde{A}&=&m^2\,\Lambda^4\,\rho^4\,\tilde{a}(g,y)\,Y^2\label{Aquad}\\
\tilde{B} &=& 2\,m^3\,\Lambda^4\rho^4\,\tilde{b}(g,y)\,|Y| \label{Bquad}\\
\tilde{C} &=& m\,\Lambda^4\rho^2\,\tilde{b}(g,y)\,|Y| \label{Cquad}
\eea
where the functions $\tilde{a},\,\tilde{b}$ are defined as
\bea
\tilde{a}(g,y)&=&(4+3g^2) + 3g\sqrt{4+g^2}\,y\label{hfunc}\\\
\tilde{b}(g,y)&=&(2+g^2)y + g\sqrt{4+g^2}\label{ffunc}
\eea
where $y\equiv Y/|Y|=\pm1$ is the sign of $Y$. From the form of Eqs.~(\ref{Aquad},\,\ref{Bquad},\,\ref{Cquad}), we see that all other factors are positive, and hence causality is determined by the signs of these dimensionless functions $\tilde{a}$ and $\tilde{b}$.

\begin{figure}[t]
\centering
\includegraphics[width=\columnwidth]{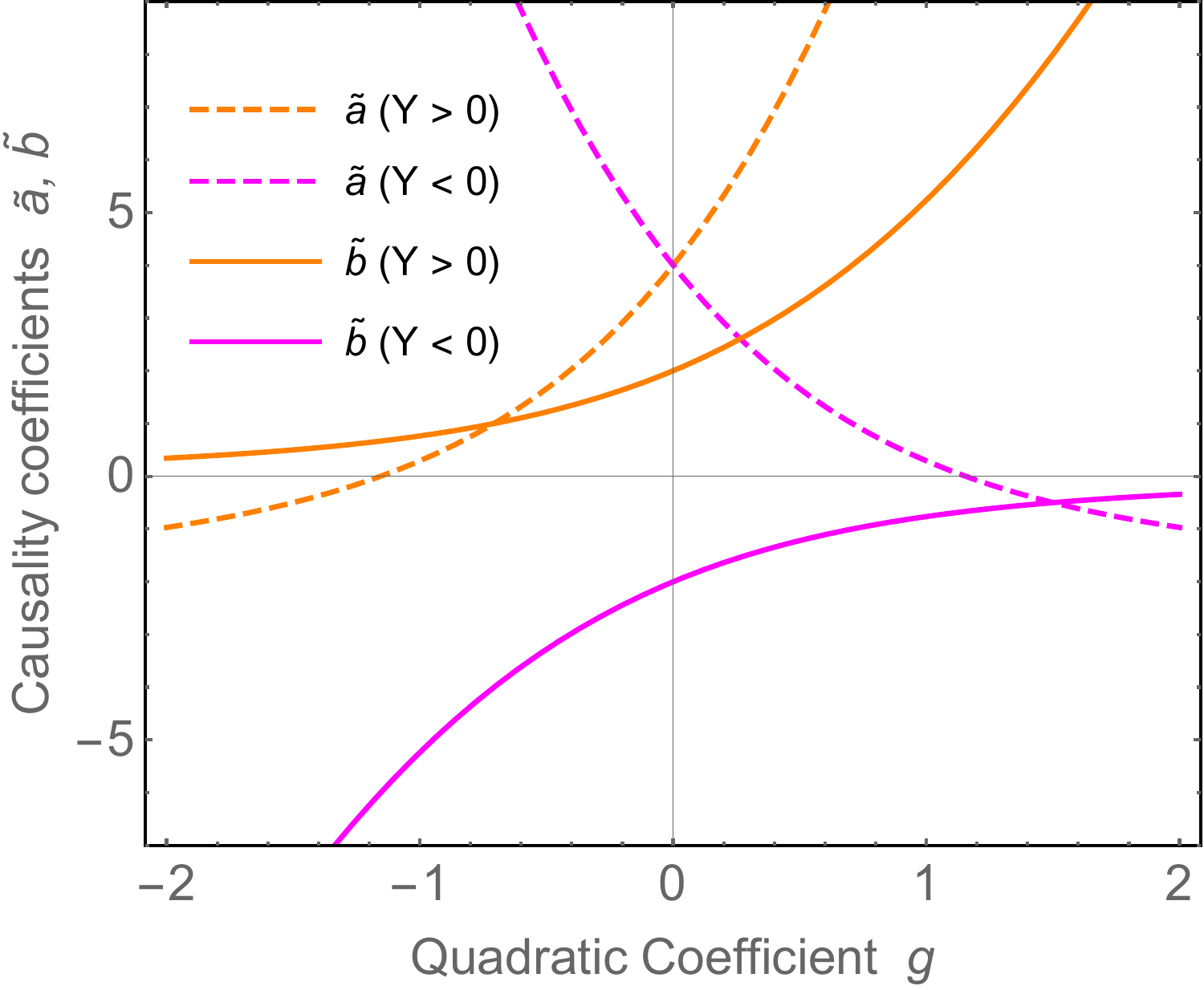} 
\caption{A plot of the functions $\tilde{a}(g,y)$ and $\tilde{b}(g,y)$ as a function of coefficient $g$; see Eqs.~(\ref{hfunc},\,\ref{ffunc}). 
Dashed curves are $\tilde{a}(g,y)$ and solid curves are $\tilde{b}(g,y)$. 
Orange curves are $y=+1$ (i.e., for $Y>0$) and magenta curves are $y=-1$ (i.e., for $Y<0$).}
\label{fig:f} 
\end{figure}

The two branches of the functions $\tilde{a}$ and $\tilde{b}$ are plotted in Figure \ref{fig:f}. Note that if $\tilde{b}>0$ on both branches it would 
help to 
allow the theory to
avoid causality problems by ensuring $\tilde{B}>0,\,\tilde{C}>0$. However, we see that for the lower branch $y=-1$ (i.e., for $Y<0$), we have $\tilde{b}<0$. Although for larger values of $g$ this function rises, it never crosses zero. 
In addition, we see that the function $\tilde{a}$ is positive for $|g|<2/\sqrt{3}$; we therefore must remain in this regime to avoid breaking the first hyperbolicity condition (A).
Altogether, since $Y\approx-(\nabla\theta)^2/(2m)<0$ in the MOND regime, we once again have that $B<0$ and the theory violates hyperbolicity. 

Interestingly, for $g>2/\sqrt{3}$, we violate {\em both} hyperbolicity conditions on the $Y<0$ (magenta) curves. This means that the effective metric alters its entire sign. In such a regime there is again some kind of causal evolution with all $A,\,B,\,C$ negative (although there are potential problems when coupling to ordinary matter due to a negative Hamiltonian for $\pf_j$). However, one can show that this is precisely the regime in which one loses the MONDian phenomenology, as the effective action then carries the wrong sign ($F=-\ac\,Y\sqrt{|Y|}$ with $\ac<0$) and the force becomes repulsive (see Eq.~(\ref{accel})). {\em So this shows that the MONDian regime is precisely at odds with causality.}

\subsection{General Analysis}\label{GenAn}

In fact we can prove the violation of hyperbolicity in the most general theory of this form. In particular, consider the most general form of $F$ for the SFDM by allowing for any arbitrary sum over terms as
\beq
F = (X+m^2|\Phi|^2)\sum_{n=0}g_n{\Lambda^{2n}\left(X+m^2|\Phi|^2\right)^{n}\over(\Lambda_c^2+|\Phi|^2)^{3n}}
\label{SuperLgenExtend}
\eeq
Note that with $g_0=1/2$, $g_1=-g/2$, $g_2=1/6$, $g_4=g_5=\ldots=0$, we recover the model studied above. Each term here carries precisely the scaling that is needed to achieve the effective action $F= -\alpha\,Y\sqrt{|Y|}$ in the MONDian limit. Also by taking $g_0>0$ we can recover CDM on large scales. 

Now what is important to notice is that inside the sum is a function of only one variable. We can express this in a general way by summarizing this as
\beq
F(X,\Phi)=(X+m^2|\Phi|^2)\,\fu\!\left(\Lambda^2(X+m^2|\Phi|^2)\over(\Lambda_c^2+|\Phi|^2)^3\right)
\eeq
where $\fu$ is some dimensionless function of its dimensionless argument
\beq
\argf\equiv{\Lambda^2(X+m^2|\Phi|^2)\over(\Lambda_c^2+|\Phi|^2)^3}
\eeq
As an example, the model studied above is the function $\fu(\argf)=1/2-\argf/2+\argf^2/6$. 
In fact we will be able to make a general proof for any $\fu$, regardless of whether or not it possesses the above regular series expansion (i.e., we can even allow for fractional powers or various other non-analytic functions in this discussion). 

With this decomposition, we can use the chain rule to readily obtain the $A$ and $C$ parameters that determine causality as
\bea
A&=&{\partial F\over\partial X}=\fu+\argf\,\fu_\argf \label{AgenZ}\\
C&=&-{\partial^2F\over\partial X^2}=-{\Lambda^2\over(\Lambda_c^2+|\Phi|^2)^3}\left(2\,\fu_\argf+\argf\,\fu_{\argf\argf}\right) \label{CgenZ}
\eea
where $\fu_\argf\equiv\partial\fu/\partial\argf$ and $\fu_{\argf\argf}\equiv\partial^2\fu/\partial\argf^2$ indicates derivates with respect to the argument $\argf$. Combined with $B=A-2XC$, we obtain all three parameters.

Now in the MOND regime we always have Eq.~(\ref{Xapprox}) and $|\Phi|=\rho\gg\Lambda_c$, this allows us to simplify $F$ as
\beq
F\approx-2\,m\,\rho^2\,Y\,\fu\left(-{2\Lambda^2\,m\,Y\over\rho^4}\right)
\label{FgenM}\eeq
Now recall that in the MOND regime we need this to become $F=-\alpha\,Y\sqrt{|Y|}$, with $\alpha>0$ to ensure attraction. So we immediately see that we need
\beq
\fu>0\,\,\,\,\,\,(\mbox{Attractive Force})
\eeq

Furthermore, Eq.~(\ref{FgenM}) shows that $F$ is acting as an effective potential for $\rho$; so we need $\rho$ to sit at its minimum. This means its derivative with respect to $\rho$ must vanish, i.e.,
\beq
{\partial F\over\partial\rho}=-4\,m\,\rho\,Y\,\fu-{16\,m^2\,\Lambda^2\,Y^2\over\rho^3}\fu_\argf=0
\label{FirstDeriv}\eeq
Since $Y\approx-(\nabla\theta)^2/(2m)<0$ in the MOND regime, the prefactor in front of $\fu$ in this equation is positive. Since we already learnt that $\fu$ must be positive for an attractive force, this equation tells us that the derivative must be positive too
\beq
\fu_\argf>0\,\,\,\,\,\,(\mbox{Attractive Force})
\eeq
Note that $\argf=-2\Lambda^2mY/\rho^4>0$ in MOND regime and hence we see from Eq.~(\ref{AgenZ}) that 
\beq
A>0
\eeq
So the first hyperbolicity condition is always obeyed whenever we have an attractive MOND force. Conversely, if we had a situation in which $A<0$, then we must have a negative value of $\fu$, and hence a negative $\alpha$ giving a repulsive force; as we saw occur in the previous subsection with $g>2/\sqrt{3}$.

Now in order to have a heavy massive $\rho$ that we can reliably integrate out, we need that the second derivative of $F$ with respect to $\rho$ is positive (if it were negative, we would have a tachyonic instability). The second derivative is readily found to be
\beq
{\partial^2F\over\partial\rho^2} = -4mY\fu+{16m^2\Lambda^2Y^2\over\rho^4}\fu_\argf -{128m^3\Lambda^4Y^3\over\rho^8}\fu_{\argf\argf}
\label{SecondDeriv}\eeq
Let us now use Eq.~(\ref{FirstDeriv}) to solve for $\fu$ in favor of $\fu_\argf$ and re-arranging, leads to
\beq
{\partial^2F\over\partial\rho^2} ={128m^2\Lambda^2Y^2\over\rho^4}\left({1\over4}\fu_\argf - {m\Lambda^2Y\over\rho^{4}}\fu_{\argf\argf}\right)
\eeq
By demanding that this is positive, so that perturbations in $\rho$ are stable, we obtain the inequality
\beq
 {m\Lambda^2Y\over\rho^{4}}\fu_{\argf\argf}<{1\over4}\fu_\argf\,\,\,\,\,\,(\mbox{Stability})
\label{stability}\eeq
Now let us see what this implies for the causality parameter $C$. In the MOND regime Eq.~(\ref{CgenZ}) is
\beq
C=-{2\Lambda^2\over\rho^6}\left(\fu_\argf-{m\Lambda^2Y\over\rho^4}\fu_{\argf\argf}\right)
\eeq
Using the stability requirement Eq.~(\ref{stability}) to bound the second term here, leads to an upper bound on $C$ of
\beq
C<-{3\Lambda^2\over2\rho^6}\fu_\argf
\label{Cbd}\eeq
Since we earlier derived that we must have $\fu_\argf>0$ in order to have an attractive force (as it is enforced by $\fu>0$), we see from Eq.~(\ref{Cbd}) that we must have $C<0$. 

Finally, we can always estimate $B\approx2m^2\rho^2C$, using $-X\approx m^2\rho^2\gg X+m^2|\Phi|^2$ in the MOND regime. Hence we obtain that $B$ must also obey
\beq
B<0
\eeq
Thus proving that the second hyperbolicity condition is broken for this entire class of models; ensuring ghost behavior of high energy perturbations (and $C<0$ means superluminality can occur in phase space in some intermediate transition regions that happen to have $X$ small, as we mentioned earlier).

\subsection{Mass Scales and Regime of Validity}

Let us discuss in some more detail, the specific parameters of the above theory. It is suggested in Ref.~\cite{Khoury} that the mass may take on the value $m\sim$\,eV. Regarding $\Lambda$ and $\Lambda_c$ it is suggested that these scales may be comparable on galactic scales, with $\Lambda_c$ smaller. However, beyond that there is some ambiguity in their values; it is suggested that their value depends sensitively on the state of the system. The proposal is that on cosmological scales, one has $\Lambda\gg m$, while on galactic scales, one has $\Lambda\ll m$. In order to achieve such a dramatic change in values, it is suggested that $\Lambda$ is an extremely sensitive function of temperature, such as $\Lambda=\Lambda_0/(1+\kappa\,(T/T_c)^{1/4})$. The idea is that in the superfluid phase, the system thermalizes, heats up, and then $\Lambda$ drops appreciably. 

If one takes this seriously (though to our knowledge, there is no concrete evidence that this actually occurs) then one might question the high energy analysis of the above subsections, where we assumed the perturbation $\varepsilon$ could be relativistic. One might think one could avoid this regime by assuming that on galactic scales the scale $\Lambda_c$, which may act as a cut off on the effective theory, takes on low values, preventing any treatment of relativistic perturbations and thus avoiding any concerns about acausality. In fact concerns of instability were already claimed in Ref.~\cite{Khoury} to be avoided due to thermal effects; but this was all done at the level of the low energy non-relativistic theory.

However, we do not think this is actually a problem with our analysis of high energy (relativistic) perturbations for the following reasons: one is always allowed to consider any classical field configurations, and study high energy perturbations above it. The fact that such a configuration may try to thermalize seems to be a separate issue. What one can say is that if one sets up a non-thermal background $\varphi_b$, as we have done here, it should not thermalize quicker than the speed of its relativistic perturbations. In fact the perturbations were shown to allow for regions of hyperbolicity breakdown (deep in MOND-like regime) and superluminality (in transition region). Hence it would be potentially problematic for the theory if thermal fluctuations were even faster and altered this analysis. This would seem to require another form of acausal behavior to explain away another. 

Moreover, if one were to insist that the cut off is always much lower than $m$, then the so-called relativistic completion is not really a full completion at all, as most reasonable Lorentz invariant effective theories should make sense to energies above the particle mass; else one is appealing to the breakdown of unitarity to save the theory, which is another potential problem.\footnote{In  Ref.~\cite{Addazi:2018ivg}, it was argued that the model of Eq.~\ref{SuperL} has a non-Wilsonian UV ``self-completion'', which is an attempt to handle the presence of higher dimensional operators, which could otherwise lead to unitarity violation in high energy scattering. If true, this would reinforce our claims, because then we could certainly perform a high energy computation. Note that the analysis of that paper does not address the issues of acausality found here.} Hence we conclude that the acausality demonstrated in this section is in fact a property of the {\em theory} itself, even if the proposed MOND regime involves somewhat different parameters. Nevertheless, a further study of relativistic perturbations in thermal states, and a more complete analysis of state space, is desirable.

\section{Other Models in the Literature}\label{Other}

In this section, we briefly discuss a few other interesting formulations of SFDM and their various features from the point of view of causality and locality.

\subsection{Higher Derivative Model}\label{Galileon}

In Ref.~\cite{Khoury:Another}, another approach was taken to extend the superfluid model of Ref.~\cite{Khoury}. The full relativistic Lagrangian there includes the terms
\bea
\mathcal{L} &=& M_{\text{Pl}}^2\left({\mathcal{R}\over2} + {\square X\over m^2}\right) \left( {1 \over 1+\chi^2} + {(\partial X)^2 \over 9m^4 a_0^2} \chi^2 \right)\nonumber\\
&&+{\Lambda^4\over n}\left(-{X\over m^2}-{1\over2}\right)^n+\ldots \label{gal}
\eea
where $a_0$ is the MOND parameter, $\mathcal{R}$ is the Ricci scalar, $M_{\text{Pl}}$ is the Planck mass, and
\beq
X = {1\over2}(\partial\Theta)^2 
\eeq
is the kinetic term of some Lorentz invariant Goldstone field $\Theta$ (in Ref.~\cite{Khoury:Another} they use notation $\mathcal{Y}\equiv-X$). 
Also, $\chi$ is another scalar field that acquires a vacuum expectation value in the MOND regime, engineered such that the model reproduces the desired MOND phenomenology. 
This model is somewhat more complicated than that of Eq.$\,$(\ref{SuperL}), noting the non-minimal coupling of 
$X$ and $\chi$ to gravity in Eq.$\,$(\ref{gal}) and the spontaneous $\mathbb{Z}_2$ symmetry breaking (and that we are writing only terms that are relevant to our discussion here).

A potential problem we wish to comment on is that the theory contains a term of the form 
\beq
\Delta\mathcal{L}=\pref\, \square X\, (\partial X)^2
\eeq
where $\pref=M_{\text{Pl}}^2\chi^2/(9\,m^6 a_0^2)>0$. In this model, $\chi$ is typically slowly varying; we will treat it as a constant here for simplicity, and focus on the dynamics of $X$. We shall expand around a background, as we did earlier, by writing $\Theta=\Theta_b+\pf$. For a simple analysis, let us take the background to be a function of time only, while allowing $\pf$ to depend on both space and time. We work to quadratic order in the perturbation $\pf$, and find the Lagrangian for these perturbations is
\beq
\Delta\mathcal{L}_2=2\pref\,\dot\Theta_b^2\!\left(\ddot\Theta_b^2+\dot\Theta_b\,\dddot\Theta_b\right)(\nabla\dot\pf)^2
\label{AltLag}\eeq
We see that the only term that appears is a term with a mixed time and spatial derivatives acting on $\pf$. Hence, whether the perturbations are well behaved or not is determined by the sign of $\ddot\Theta_b^2+\dot\Theta_b\,\dddot\Theta_b$; if it is positive, then it is standard, while if it is negative, then it violates hyperbolicity and is a ghost. 

In the MOND regime, one has $\Theta=m\,t+\theta$, where $\theta$ is slowly changing. This means that if we say that $\theta$ changes with a characteristic rate $\Gamma$, then $\Gamma\ll m$. This allows us to estimate $\ddot\Theta^2=\ddot\theta^2\sim\Gamma^4\theta$ and $\dot\Theta\,\dddot\Theta\approx m\,\dddot\theta\sim m\,\Gamma^3\theta$, and hence this second term dominates. This means the sign of the Lagrangian Eq.~(\ref{AltLag}) is determined by the sign of $\dddot\theta$. So if $\dddot\theta<0$, then there is a breakdown of hyperbolicity. Note that this is evidently not a heavy ghost mode, as it arises from a massless $\theta$ mode in a second order in time equation of motion. Instead it is a light ghost mode, and so cannot be easily ignored. To our knowledge, there is no consistent way to restrict phase space to always forbid $\dddot\theta<0$, and therefore this hyperbolicity breakdown can occur within the theory. By coupling $\pf$ to other degrees of freedom, there can be a ghost-like instability due to the corresponding inverted Hamiltonian.

One may wish to first rewrite Eq.~(\ref{gal}) in the Einstein frame, which will lead to some potential for the fields 
$V=V(X,\,\chi)$ in the Lagrangian with the canonical Einstein-Hilbert term. However, this rewriting of (\ref{gal}) is anticipated to necessarily still contain terms 
$\sim \square X\,(\partial X)^2$; whatever Weyl transformation and redefinition we effect to make the gravitational kinetic term canonical will not disrupt the presence of this term. 
This suggests that while such a theory may be an alternative to the model in Eq.\,(\ref{SuperL}), it includes further problems. 

A second potential problem we would like to comment on, arises from the term on the second line of Eq.~(\ref{gal}). For integer powers $n$, we can expand this around the vacuum, giving the leading terms
\beq
F={(-1)^{n+1}\Lambda^4\over2^{n-1}m^2}X+{(-1)^{n+1}\Lambda^4\over 2^{n-1}(n-1)\,m^4}X^2+\ldots
\eeq
We see that if $n$ is even (which includes the $n=2$ case that is highlighted in Ref.~\cite{Khoury:Another}), then the first term here is a ghost. While if $n$ is odd, then the second term here violates the condition for subluminality, i.e., the leading interaction has $-F''<0$. Finally, if $n$ is non-integer (which includes the $n=5/2$ case that is also highlighted in Ref.~\cite{Khoury:Another}), then this term in the second line of Eq.~(\ref{gal}) would not be well defined for $X/m^2+1/2>0$. To avoid this problem in all of phase space, one would require an absolute value in order to avoid problems when studying fluctuations around the vacuum. Note that this is very different to the DBI models studied earlier, due to a different overall sign inside the square root in Eq.~(\ref{DBI}), and so those are well behaved around the vacuum. (As pointed out in Ref.~\cite{Khoury:Another}, the MOND regime obeys the inequality $X/m^2+1/2<0$, but we are examining the plausibility of embedding such theories into microphysics, for which one cannot normally restrict the phase space in this way.) If one were to then include an absolute value, the Lagrangian would have a singular structure and would fail basic locality tests, such as cluster decomposition.

\subsection{Inverse Operators }

In Ref.~\cite{Khoury:USFDS} another interesting class of models was developed. In this case the authors needed to include a term of the form 
\beq
\Delta\mathcal{L}\propto -{\Psi_1^*\Psi_2+\Psi_2^*\Psi_1\over|\Psi_1||\Psi_2|}\label{P12}
\eeq
in order to obtain the desired MONDian phenomenology. This is certainly a peculiar interaction. Since it involves inverse operators, it is anticipated to fail the cluster decomposition principle for a basic notion of locality \cite{Weinberg}. We can be more concrete about this as follows: Usually one should form an effective field theory in order to track its regime of validity. We can alter this theory in the following way to define a kind of effective theory
\beq
\Delta\mathcal{L}\propto -{\Psi_1^*\Psi_2+\Psi_2^*\Psi_1\over\sqrt{(|\Psi_1|^2+\Lambda^2)(|\Psi_2|^2+\Lambda^2)}}
\eeq
where we have introduced a new energy scale $\Lambda$.
This theory now permits a regular series expansion around the vacuum $\Psi_1=\Psi_2=0$. The cutoff of this effective theory is now $\sim\Lambda$. In order to recover the model with MONDian dynamics of Eq.$\,$(\ref{P12}) one would need to send the cutoff $\Lambda\to0$. This reflects the idea that it carries a kind of non-locality as follows: For non-zero $\Lambda$, we have a cutoff length scale $L_c\sim 1/\Lambda$. In some sense, any effective theory is not strictly local because one cannot localize particles on scales smaller than this scale $L_c$. If one now sends $\Lambda\to0$, then the cutoff length scale $L_c\to\infty$, which means that one cannot define locality on any finite scale in the usual sense.

\subsection{Split Role Model}\label{TwoField} 

Finally, we mention an attempt to reconcile what Ref.~\cite{Mistele} describes as a tension between a superfluid in equilibrium and a significant superfluid energy density by splitting these two roles between two different fields. 
This paper studied another K-essence theory of the form 
\bea
\mathcal{L} &=& {1\over2}\left[X_{\rho_-} + \rho_-^2\left(X_- - m^2\right)\right] - {\lambda_4 \over 4} \rho_-^4 \nonumber \\
&&+ F\left(X_++X_--m^2\right)-\lambda\theta_+\rhob \label{split}
\eea
where $X_{\pm} \equiv \partial^\mu \theta_{\pm} \partial_\mu \theta_{\pm}$ for two real scalars $\theta_{\pm}$, $\rho_-$ is a canonical real scalar component and
$F(X) = 2\,\Lambda\,X \sqrt{|X|}$/3.
The shift symmetry of $\theta_-$ is identified as corresponding to the chemical potential of the superfluid in the shift $\dot{\theta}_- \to \dot{\theta}_- + \mu$. Only $\theta_+$ is coupled to baryonic matter $\rhob$, separating these two roles of the superfluid (we note that the energy density $\rhob$ is not Lorentz invariant, but it was the choice of Ref.~\cite{Mistele}).

By perturbing around a background solution, the sound speed of $\theta_+$ propagations were found to be superluminal,
\beq
	c_+^2 = 1+ \gamma^2 \geq 1
\eeq
where $\gamma$ is the cosine of the angle between the gradients of the background field and the perturbation. However, by modifying the theory to include mixing between $\theta_\pm$
\bea
	&&F\left(X_++X_--m^2\right) \nonumber\\
	&&\,\,\,\,\,\,\,\to F\left(X_+ +X_- - m^2 + \Cfix\left[\partial^\mu \theta_+ \partial_\mu \theta_-\right]^2\right) \label{mod}
\eea
Ref.~\cite{Mistele} found the sound speed becomes 
\beq
	c_+^2 = {1+\gamma^2 \over 1+\Cfix \mu^2} \label{subluminal}
\eeq
While the sound speeds are still superluminal in some region of parameter space, the author points out one can choose $\Cfix$ and truncate to some parts of phase space where the sound speeds are always subluminal. 

However, we would like to emphasize that this is only addressing the sound speeds of the theory, which are collective excitations of the superfluid. This does not address the issue of the structure of propagation of elementary high energy particles on the superfluid background; these are in fact acausal, as shown earlier in our paper. 

Furthermore, this theory prevents standard cluster decomposition as it is a singular action. So modifying (\ref{split}) with (\ref{mod}) fails to fix the major problems of this class of models. Ref.$\,$\cite{Mistele} includes an appendix which describes a way to lift Eq.$\,$(\ref{split}) to a smooth Lagrangian, but the form of the theory arrived at by this process is simply of the form of a multi-field version of Eq.$\,$(\ref{SuperL}). So it again fails the conditions for global hyperbolicity (in MOND regime) and subluminality (in transition regime) as we have shown in Sec.$\,$(\ref{SuperfluidTest}).

\section{Conclusions}\label{Conclusions}

We have studied several models that exhibit MONDian phenomenology on galactic scales. We focussed on the so-called superfluid dark matter models that also relax to standard CDM on larger scales. We found that, while such an approach provides a novel dark matter model that can conform to the observed galactic velocity dispersion, each Lorentz invariant model contained some form of acausal behavior that could not easily be resolved.  

We found that the most basic types of Lorentz invariant models in Eqs.~(\ref{nonsmth},\,\ref{smth}), which exhibit the MONDian behavior, violate the condition (C) for subluminality of high energy perturbations, Eq.$\,$(\ref{C}). One might have hoped that the superfluid dark matter models, which seem to be built on more reasonable Lagrangians, would be better behaved. However, the superfluid dark matter models share their own potential problems. The primary model we considered was that of Eq.$\,$(\ref{SuperL}). In Section \ref{SuperfluidTest}, we began by generalizing the causality conditions to a complex field. We then applied these to this class of models. We found that this theory in fact violates condition (B) for hyperbolicity, Eq.$\,$(\ref{B}), leading to ghost behavior of relativistic perturbations. 
The corresponding Hamiltonian of perturbations indeed has a flipped sign of its temporal derivative terms. This is a breakdown in some forms of causal propagation.
Formally condition (C) was violated too, although the speed of propagation becomes undefined in this regime. Nevertheless, it can lead to superluminal behavior in parts of phase space in intermediate transition regions from CDM to MOND (such as a region of space that happens to be semi-relativistic). We then generalized the basic model, establishing a general proof that they all violate these basic conditions for causal propagation.

Altogether we note that in the transition from CDM towards MOND, the sound speed can become large, leading to superluminality and a regular form of acausality. Then deeper into the MOND regime, hyperbolicity is broken. This means that Lorentz invariant effective theory is failing even before we enter the MOND regime, and so this regime does not appear to be present with any standard effective theory.

In this class of models, there is a subtlety associated with the scales and the regime of validity of the effective theory. Normally a consistent Lorentz invariant effective theory has a cut off well above the particle mass, allowing for the study of high energy (relativistic) perturbations, as we have done here. However, one might try to (i) lower the cut off or (ii) appeal to thermal corrections to avoid potential problems. Either option can be problematic, as we discuss:

(i) Lowering the cut off to below the particle mass means severely restricting the use of the Lorentz invariant completion and thereby appealing to very low energy new physics. Furthermore if the cut off is so low, then for all intents and purposes, these purported Lorentz invariant theories do not exist as reliable theories in the first place. Indeed as we have shown, the Lorentz invariant formulations suffer various forms of acausality. So in that sense, precisely because of this work, they don’t possess sensible Lorentz invariant completions. Finally, one might consider higher derivative corrections to the effective action. Since again we only need moderately relativistic perturbations, then if these terms were important it would mean an extremely low cut off and likely to lead to other ghosts through the higher derivative terms becoming active at such low scales. Altogether it seems unlikely that higher derivatives could cure the problem. Let us furthermore note that the present theories fail some of the “sign theorems” of the literature, and it is generally agreed that higher derivatives cannot cure this problem.

(ii) A thermalization process that avoids the existing non-thermal (relativistic) ghosts and superluminality seems difficult to achieve in a consistent way, without producing new problems in the underlying quantum field theory. Let us note that the superfluid regime is really already the low temperature phase of the theory. So it does not seem at all clear that one can appeal to finite temperature corrections to alleviate the problems of the theory. Again perturbations in the theory that are relativistic are either superluminal or violate hyperbolicity. Appealing to thermal effects to avoid this seems rather unlikely. All these topics deserve further consideration. 

In Section \ref{Other} we commented on some other related models. In particular, we mentioned the model in Eq.$\,$(\ref{gal}), which we see contains higher derivative terms, which we found can exhibit a kind of hyperbolicity breakdown, and other terms that can also feature ghosts or superluminality. We stress that the problem is not that the low energy non-relativistic effective theory carries fraction powers, but that some of the corresponding Lorentz invariant completions do too; this would cause problems with cluster decomposition, but smoothing the theories to avoid this did not alleviate the superluminality breakdown.
We also considered a theory with inverse interaction operators, as in Eq.$\,$(\ref{P12}). While the theory without this term is taken to be standard, this interaction term violates a basic notion of locality as the cluster decomposition principle and is not part of a standard consistent effective field theory. Again the non-local character seems essential to achieve the MONDian phenomenology.
Finally we mentioned a two-field model that splits the roles of dark matter and the MONDian force carrier, Eq.$\,$(\ref{split}). In this work it was pointed out that  superluminality of the sound speeds could be cured by deforming the theory. However, this did not at all address the issue of acausality in the high energy (relativistic) perturbations. 

In summary, if one can accept such novel theories, they can provide various desired phenomenological results on galactic and cosmological scales. This is very interesting at the phenomenological level. However, with the forms of acausality, including ghost instabilities, superluminality, and other kinds of non-locality established in our work, it brings into doubt whether one can embed such effective theories within microphysics. Furthermore, they would resist having a (standard) Lorentz invariant UV completion. On the other hand, an observational confirmation of these models would therefore be a spectacular discovery of completely new behavior in nature, never encountered in standard quantum field theory, and would have profound consequences.

\section*{Acknowledgments}
M.~P.~H is supported in part by National Science Foundation grant PHY-2013953.



\begin{thebibliography}{}

%
\bibitem{Planck2018}
N.~Aghanim \textit{et al.} [Planck],
``Planck 2018 results. VI. Cosmological parameters,''
Astron. Astrophys. \textbf{641}, A6 (2020)
[arXiv:1807.06209 [astro-ph.CO]].


%
\bibitem{McGaugh:2000}
S.~S.~McGaugh, J.~M.~Schombert, G.~D.~Bothun and W.~J.~G.~de Blok,
``The Baryonic Tully-Fisher relation,''
Astrophys. J. Lett. \textbf{533}, L99-L102 (2000)
[arXiv:astro-ph/0003001 [astro-ph]].
%
\bibitem{McGaugh:2011}
S.~McGaugh,
``The Baryonic Tully-Fisher Relation of Gas Rich Galaxies as a Test of LCDM and MOND,''
Astron. J. \textbf{143}, 40 (2012)
[arXiv:1107.2934 [astro-ph.CO]].


%
\bibitem{Vogelsberger}
M.~Vogelsberger, S.~Genel, V.~Springel, P.~Torrey, D.~Sijacki, D.~Xu, G.~F.~Snyder, D.~Nelson and L.~Hernquist,
``Introducing the Illustris Project: Simulating the coevolution of dark and visible matter in the Universe,''
Mon. Not. Roy. Astron. Soc. \textbf{444}, no.2, 1518-1547 (2014)
[arXiv:1405.2921 [astro-ph.CO]].


%
\bibitem{Boylan-Kolchin}
M.~Boylan-Kolchin, J.~S.~Bullock and M.~Kaplinghat,
``The Milky Way's bright satellites as an apparent failure of LCDM,''
Mon. Not. Roy. Astron. Soc. \textbf{422}, 1203-1218 (2012)
[arXiv:1111.2048 [astro-ph.CO]].
%
\bibitem{Pawlowski:2013kpa}
M.~S.~Pawlowski, P.~Kroupa and H.~Jerjen,
``Dwarf Galaxy Planes: the discovery of symmetric structures in the Local Group,''
Mon. Not. Roy. Astron. Soc. \textbf{435}, 1928 (2013)
[arXiv:1307.6210 [astro-ph.CO]].
%
\bibitem{Ibata}
R.~A.~Ibata, G.~F.~Lewis, A.~R.~Conn, M.~J.~Irwin, A.~W.~McConnachie, S.~C.~Chapman, M.~L.~Collins, M.~Fardal, A.~M.~N.~Ferguson and N.~G.~Ibata, \textit{et al.}
``A Vast Thin Plane of Co-rotating Dwarf Galaxies Orbiting the Andromeda Galaxy,''
Nature \textbf{493}, 62-65 (2013)
[arXiv:1301.0446 [astro-ph.CO]].
%
\bibitem{deBlok:2009sp}
W.~J.~G.~de Blok,
``The Core-Cusp Problem,''
Adv. Astron. \textbf{2010}, 789293 (2010)
[arXiv:0910.3538 [astro-ph.CO]].


%
\bibitem{Milgrom:1983ca}
M.~Milgrom,
``A Modification of the Newtonian dynamics as a possible alternative to the hidden mass hypothesis,''
Astrophys. J. \textbf{270}, 365-370 (1983).
%
\bibitem{Milgrom:1983pn}
M.~Milgrom,
``A Modification of the Newtonian dynamics: Implications for galaxies,''
Astrophys. J. \textbf{270}, 371-383 (1983)
%
\bibitem{Milgrom:1983zz}
M.~Milgrom,
``A modification of the Newtonian dynamics: implications for galaxy systems,''
Astrophys. J. \textbf{270}, 384-389 (1983)
%
\bibitem{Bekenstein:1984tv}
J.~Bekenstein and M.~Milgrom,
``Does the missing mass problem signal the breakdown of Newtonian gravity?,''
Astrophys. J. \textbf{286}, 7-14 (1984)


%
\bibitem{Weinberg:1964ew} 
  S.~Weinberg,
  ``Photons and Gravitons in s Matrix Theory: Derivation of Charge Conservation and Equality of Gravitational and Inertial Mass,''
  Phys.\ Rev.\  {\bf 135}, B1049 (1964).
%
  \bibitem{Weinberg:1965rz}
S.~Weinberg,
``Photons and gravitons in perturbation theory: Derivation of Maxwell's and Einstein's equations,''
Phys. Rev. \textbf{138}, B988-B1002 (1965).
%
\bibitem{Deser:1969wk}
S.~Deser,
``Selfinteraction and gauge invariance,''
Gen. Rel. Grav. \textbf{1}, 9-18 (1970)
[arXiv:gr-qc/0411023 [gr-qc]].
%
\bibitem{Feynman}
R.~P.~Feynman, F.~B.~Morinigo, W.~G.~Wagner, B.~Hatfield, 
``Feynman Lectures on Gravitation," 
Addison-Wesley (1995). 
%
\bibitem{Khoury:2013oqa}
J.~Khoury, G.~E.~J.~Miller and A.~J.~Tolley,
``On the Origin of Gravitational Lorentz Covariance,''
Class. Quant. Grav. \textbf{31}, 135011 (2014)
[arXiv:1305.0822 [hep-th]].
%
\bibitem{Hertzberg:2016djj}
M.~P.~Hertzberg,
``Constraints on Gravitation from Causality and Quantum Consistency,''
Adv. High Energy Phys. \textbf{2018}, 2657325 (2018)
[arXiv:1610.03065 [hep-th]].
%
\bibitem{Hertzberg:2017abn}
M.~P.~Hertzberg and M.~Sandora,
``General Relativity from Causality,''
JHEP \textbf{09}, 119 (2017)
[arXiv:1702.07720 [hep-th]].
%
\bibitem{Hertzberg:2017nzl}
M.~P.~Hertzberg and M.~Sandora,
``Special Relativity from Soft Gravitons,''
Phys. Rev. D \textbf{96}, no.8, 084048 (2017)
[arXiv:1704.05071 [hep-th]].
%
\bibitem{Hertzberg:2020yzl}
M.~P.~Hertzberg, J.~A.~Litterer and M.~Sandora,
``Symmetries from locality. II. Gravitation and Lorentz boosts,''
Phys. Rev. D \textbf{102}, no.2, 025023 (2020)
[arXiv:2005.01744 [hep-th]].
%
\bibitem{Hertzberg:2020gxu}
M.~P.~Hertzberg and J.~A.~Litterer,
``Symmetries from locality. III. Massless spin-2 gravitons and time translations,''
Phys. Rev. D \textbf{102}, no.8, 085007 (2020)
[arXiv:2008.06510 [hep-th]].

%
\bibitem{Bekenstein:2004ne}
J.~D.~Bekenstein,
``Relativistic gravitation theory for the MOND paradigm,''
Phys. Rev. D \textbf{70}, 083509 (2004)
[erratum: Phys. Rev. D \textbf{71}, 069901 (2005)]
[arXiv:astro-ph/0403694 [astro-ph]].
%
\bibitem{Bruneton:2008fk}
J.~P.~Bruneton, S.~Liberati, L.~Sindoni and B.~Famaey,
``Reconciling MOND and dark matter?,''
JCAP \textbf{03}, 021 (2009)
[arXiv:0811.3143 [astro-ph]].
%
\bibitem{Blanchet:2006yt}
L.~Blanchet,
``Gravitational polarization and the phenomenology of MOND,''
Class. Quant. Grav. \textbf{24}, 3529-3540 (2007)
[arXiv:astro-ph/0605637 [astro-ph]].
%
\bibitem{Bernard:2014psa}
L.~Bernard and L.~Blanchet,
``Phenomenology of Dark Matter via a Bimetric Extension of General Relativity,''
Phys. Rev. D \textbf{91}, no.10, 103536 (2015)
[arXiv:1410.7708 [astro-ph.CO]].
%
\bibitem{Skordis:2020eui}
C.~Skordis and T.~Zlosnik,
``A new relativistic theory for Modified Newtonian Dynamics,''
[arXiv:2007.00082 [astro-ph.CO]].


%
\bibitem{Calmet:2017voc}
X.~Calmet and I.~Kuntz,
``What is modified gravity and how to differentiate it from particle dark matter?,''
Eur. Phys. J. C \textbf{77}, no.2, 132 (2017)
[arXiv:1702.03832 [gr-qc]].


%
\bibitem{Sanders}
R.~H.~Sanders and S.~S.~McGaugh,
``Modified Newtonian dynamics as an alternative to dark matter,''
Ann. Rev. Astron. Astrophys. \textbf{40}, 263-317 (2002)
[arXiv:astro-ph/0204521 [astro-ph]].
%
\bibitem{Famaey:2011kh}
B.~Famaey and S.~McGaugh,
``Modified Newtonian Dynamics (MOND): Observational Phenomenology and Relativistic Extensions,''
Living Rev. Rel. \textbf{15}, 10 (2012)
[arXiv:1112.3960 [astro-ph.CO]].
%
\bibitem{Zhao}
H.~Zhao, B.~Famaey, F.~L\"ughausen and P.~Kroupa,
``Local Group timing in Milgromian dynamics. A past Milky Way-Andromeda encounter at z\ensuremath{>}0.8,''
Astron. Astrophys. \textbf{557}, L3 (2013)
[arXiv:1306.6628 [astro-ph.GA]].
%
\bibitem{Skordis}
C.~Skordis, D.~F.~Mota, P.~G.~Ferreira and C.~Boehm,
``Large Scale Structure in Bekenstein's theory of relativistic Modified Newtonian Dynamics,''
Phys. Rev. Lett. \textbf{96}, 011301 (2006)
[arXiv:astro-ph/0505519 [astro-ph]].
%
\bibitem{Dodelson}
S.~Dodelson,
``The Real Problem with MOND,''
Int. J. Mod. Phys. D \textbf{20}, 2749-2753 (2011)
[arXiv:1112.1320 [astro-ph.CO]].
%
\bibitem{Zhao:2005xk}
H.~Zhao,
``Modified Newtonian Dynamics: Success and problem on globular cluster scale,''
[arXiv:astro-ph/0508635 [astro-ph]].



%
\bibitem{Berezhiani:2015pia}
L.~Berezhiani and J.~Khoury,
``Dark Matter Superfluidity and Galactic Dynamics,''
Phys. Lett. B \textbf{753}, 639-643 (2016)
[arXiv:1506.07877 [astro-ph.CO]].
%
\bibitem{Khoury}
L.~Berezhiani and J.~Khoury,
``Theory of dark matter superfluidity,''
Phys. Rev. D \textbf{92}, 103510 (2015)
[arXiv:1507.01019 [astro-ph.CO]].
%
\bibitem{Khoury:Another}
J.~Khoury,
``Another Path for the Emergence of Modified Galactic Dynamics from Dark Matter Superfluidity,''
Phys. Rev. D \textbf{93}, no.10, 103533 (2016)
[arXiv:1602.05961 [astro-ph.CO]].
%
\bibitem{Khoury:USFDS}
E.~G.~M.~Ferreira, G.~Franzmann, J.~Khoury and R.~Brandenberger,
``Unified Superfluid Dark Sector,''
JCAP \textbf{08}, 027 (2019)
[arXiv:1810.09474 [astro-ph.CO]].





%
\bibitem{Aharonov:1969vu}
Y.~Aharonov, A.~Komar and L.~Susskind,
``Superluminal behavior, causality, and instability,''
Phys. Rev. \textbf{182}, 1400-1403 (1969)
%
\bibitem{Arkani-Hamed}
A. Adams, N. Arkani-Hamed, S. Dubovsky, A. Nicolis and R. Rattazzi,
``Causality, analyticity and an IR obstruction to UV completion,''
JHEP \textbf{10}, 014 (2006)
[arXiv:hep-th/0602178 [hep-th]].


%
\bibitem{Bruneton:2006gf}
J.~P.~Bruneton,
``On causality and superluminal behavior in classical field theories: Applications to k-essence theories and MOND-like theories of gravity,''
Phys. Rev. D \textbf{75}, 085013 (2007)
[arXiv:gr-qc/0607055 [gr-qc]].
%
\bibitem{Bruneton:2007si}
J.~P.~Bruneton and G.~Esposito-Farese,
``Field-theoretical formulations of MOND-like gravity,''
Phys. Rev. D \textbf{76}, 124012 (2007)
[erratum: Phys. Rev. D \textbf{76}, 129902 (2007)]
[arXiv:0705.4043 [gr-qc]].

%
\bibitem{Waldtextbook}
R.~M.~Wald, 
``General Relativity," 
The University of Chicago Press (1984)


%
\bibitem{Garriga:1999vw}
J.~Garriga and V.~F.~Mukhanov,
``Perturbations in k-inflation,''
Phys. Lett. B \textbf{458}, 219-225 (1999)
[arXiv:hep-th/9904176 [hep-th]].
%
\bibitem{ArmendarizPicon:1999rj}
C.~Armendariz-Picon, T.~Damour and V.~F.~Mukhanov,
``k - inflation,''
Phys. Lett. B \textbf{458}, 209-218 (1999)
[arXiv:hep-th/9904075 [hep-th]].
%
\bibitem{Rendall:2005fv}
A.~D.~Rendall,
``Dynamics of k-essence,''
Class. Quant. Grav. \textbf{23}, 1557-1570 (2006)
[arXiv:gr-qc/0511158 [gr-qc]].


%
\bibitem{Mistele}
T. Mistele,
``Three problems of superfluid dark matter and their solution,''
JCAP \textbf{2021}, 025 (2021) [arXiv:2009.03003v3 [gr-qc]].


%
\bibitem{Babichev:2007dw}
E.~Babichev, V.~Mukhanov and A.~Vikman,
``k-Essence, superluminal propagation, causality and emergent geometry,''
JHEP \textbf{02}, 101 (2008)
[arXiv:0708.0561 [hep-th]].


%
\bibitem{Silverstein:2003hf}
E.~Silverstein and D.~Tong,
``Scalar speed limits and cosmology: Acceleration from D-cceleration,''
Phys. Rev. D \textbf{70}, 103505 (2004)
[arXiv:hep-th/0310221 [hep-th]].
%
\bibitem{Alishahiha:2004eh}
M.~Alishahiha, E.~Silverstein and D.~Tong,
``DBI in the sky,''
Phys. Rev. D \textbf{70}, 123505 (2004)
[arXiv:hep-th/0404084 [hep-th]].




%
\bibitem{Weinberg}
S. Weinberg,
``The Quantum Theory of Fields, Vol. 1,''
Cambridge University Press (1995).


%
\bibitem{Khoury:2014tka}
J.~Khoury,
``Alternative to particle dark matter,''
Phys. Rev. D \textbf{91}, no.2, 024022 (2015)
[arXiv:1409.0012 [hep-th]].


%
\bibitem{Boyarsky:2008xj}
A.~Boyarsky, J.~Lesgourgues, O.~Ruchayskiy and M.~Viel,
``Lyman-alpha constraints on warm and on warm-plus-cold dark matter models,''
JCAP \textbf{05}, 012 (2009)
[arXiv:0812.0010 [astro-ph]].
%
\bibitem{Viel:2005qj}
M.~Viel, J.~Lesgourgues, M.~G.~Haehnelt, S.~Matarrese and A.~Riotto,
``Constraining warm dark matter candidates including sterile neutrinos and light gravitinos with WMAP and the Lyman-alpha forest,''
Phys. Rev. D \textbf{71}, 063534 (2005)
[arXiv:astro-ph/0501562 [astro-ph]].
%
\bibitem{Viel:2013fqw}
M.~Viel, G.~D.~Becker, J.~S.~Bolton and M.~G.~Haehnelt,
``Warm dark matter as a solution to the small scale crisis: New constraints from high redshift Lyman-\ensuremath{\alpha} forest data,''
Phys. Rev. D \textbf{88}, 043502 (2013)
[arXiv:1306.2314 [astro-ph.CO]].
%
\bibitem{Schutz:2020jox}
K.~Schutz,
``Subhalo mass function and ultralight bosonic dark matter,''
Phys. Rev. D \textbf{101}, no.12, 123026 (2020)
[arXiv:2001.05503 [astro-ph.CO]].
%
\bibitem{Benito:2020avv}
M.~Benito, J.~C.~Criado, G.~H\"utsi, M.~Raidal and H.~Veerm\"ae,
``Implications of Milky Way substructures for the nature of dark matter,''
Phys. Rev. D \textbf{101}, no.10, 103023 (2020)
[arXiv:2001.11013 [astro-ph.CO]].


%
\bibitem{Deng:2018jjz}
H.~Deng, M.~P.~Hertzberg, M.~H.~Namjoo and A.~Masoumi,
``Can Light Dark Matter Solve the Core-Cusp Problem?,''
Phys. Rev. D \textbf{98}, no.2, 023513 (2018)
[arXiv:1804.05921 [astro-ph.CO]].


\bibitem{Addazi:2018ivg}
A.~Addazi and A.~Marcian\`o,
``UV self-completion of a theory of Superfluid Dark Matter,''
Eur. Phys. J. C \textbf{79}, no.4, 354 (2019)
[arXiv:1801.04083 [hep-th]].






\end{thebibliography}
\end{document}